\documentclass[twocolumn]{aastex62}
\pdfoutput=1 %for arXiv submission
\usepackage{amsmath,amstext}
\usepackage[T1]{fontenc}
\usepackage{apjfonts} 
\usepackage[figure,figure*]{hypcap}
\usepackage{subfigure}
\usepackage{needspace}

\shorttitle{The Re-Acceleration of the Shock Wave in the Radio Remnant of SN 1987A}
\begin{document}

\title{The Re-Acceleration of the Shock Wave in the Radio Remnant of SN 1987A}

\author{\textbf{Y. Cendes}}
\affiliation{Dunlap Institute for Astronomy and Astrophysics University of Toronto, Toronto, ON M5S 3H4, Canada}
\affiliation{Leiden Observatory, PO Box 9513, 2300 RA Leiden, The Netherlands}
%\affiliation{Affiliation 2}
\author{B. M. Gaensler}
\affiliation{Dunlap Institute for Astronomy and Astrophysics University of Toronto, Toronto, ON M5S 3H4, Canada}
%\affiliation{Affiliation 3}
\author{C.-Y. Ng}
\affiliation{Department of Physics, The University of Hong Kong, Pokfulam Road, Hong Kong}
\author{G. Zanardo}
\affiliation{International Centre for Radio Astronomy Research (ICRAR) --
The University of Western Australia, Crawley, WA 6009, Australia}
\author{L. Staveley-Smith}
\affiliation{International Centre for Radio Astronomy Research (ICRAR) --
The University of Western Australia, Crawley, WA 6009, Australia}
\author{A. K. Tzioumis}
\affiliation{CSIRO Astronomy and Space Science, Australia Telescope National Facility, Marsfield, NSW 1710, Australia}

\begin{abstract}
We report on updated radio imaging observations of the radio remnant of Supernova 1987A (SN 1987A) at 9 GHz, taken with the Australia Telescope Compact Array (ATCA), covering a 25-year period (1992-2017).  We use Fourier modeling of the supernova remnant to model its morphology, using both a torus model and a ring model, and find both models show an increasing flux density, and have shown a continuing expansion of the remnant.  As found in previous studies, we find the torus model most accurately fits our data, and has shown a change in the remnant expansion at Day 9,300 $\pm$210 from 2,300 $\pm$200 km/s to 3,610 $\pm$240 km/s.  We have also seen an increase in brightness in the western lobe of the remnant, although the eastern lobe is still the dominant source of emission, unlike what has been observed at contemporary optical and X-ray wavelengths.  We expect to observe a reversal in this asymmetry by the year $\sim$2020, and note the south-eastern side of the remnant is now beginning to fade, as has  also been seen in optical and X-ray data.  Our data indicate that high-latitude emission has been present in the remnant from the earliest stages of the shockwave interacting with the equatorial ring around Day 5,000.  However, we find the emission has become increasingly dominated by the low-lying regions by Day 9,300, overlapping with the regions of X-ray emission.  We conclude that the shockwave is now leaving the equatorial ring, exiting first from the south-east region of the remnant, and is re-accelerating as it begins to interact with the circumstellar medium beyond the dense inner ring.
\end{abstract}

\section{Introduction}

Supernova 1987A (SN 1987A) is the closest observed supernova to Earth since the invention of the telescope.  The initial supernova event was observed on February 23, 1987, and since then monitoring of the supernova remnant (SNR) at multiple wavelengths has provided crucial information in understanding the remnant's evolution \citep{McCray2016}.  The SNR has been shown to evolve on a time scale of months to years, and its relatively near distance in the Large Magellanic Cloud ($\sim50$ kpc) has allowed for many details in the structure to be visible.  

The progenitor star for SN 1987A, Sanduleak -69$^{\circ}$202, was surrounded by an unusual structure of material, believed to be emitted by the progenitor before the supernova explosion.  The most striking optical feature consists of two outer rings forming an hourglass structure and one dominant equatorial ring (ER) \citep{Burrows1995}.  The origin of these rings is unclear, but it is thought they could originate from a binary merger some $\sim$20,000 years ago \citep{Urushibata2018, Menon2017, Morris2009,Morris2007}, or perhaps a fast-rotating progenitor star \citep{Chita2008}.  These rings became visible during the ionizing flash of photons released by the supernova event \citep{Burrows1995}, and the ER is located on the shock front boundary between the HII region of the red supergiant (RSG) progenitor and the RSG free wind \citep{Chevalier1995}.  The density in the polar directions is thought to be somewhat lower, and little is known about the region beyond the ER \citep{Chevalier1995,Mattila:2010}.

The first radio emission from SN 1987A was detected two days after the supernova event (February 25, 1987) by the Molonglo Observatory Synthesis Telescope (MOST) \citep{Turtle1987}.  This emission reached a peak four days after the explosion, then faded below a $3\sigma$ detection limit some $\sim200$ days after the explosion \citep{Ball1995}.  Emission was then detected again $\sim$1,200 days after the supernova event (mid-1990), both with MOST and the Australia Telescope Compact Array \citep[ATCA;][]{Ball1995,SS1992}.  Since then, constant monitoring in the 9 GHz band shows that the radio remnant has steadily increased in radio emission, thought to originate from synchrotron emission as electrons are accelerated by the expanding shockwaves from the original supernova explosion \citep{Ball1992}.  Continuous multi-wavelength monitoring suggests that the shockwave reached the optical inner ring around day 5,500-6,500, which corresponded with an increase in observed radio luminosity \citep{Zanardo2010}.

Recent data at both visible and X-ray wavelengths suggest that the SNR is now reaching the end of its current phase of passing through the ER.  \citet{Fransson2015} reported that the most dense clumps of the emission region have been fading since Day $\sim$8,000, with diffuse emission and hot spots appearing outside the optical ring around Day $\sim$9,500.  They predicted that the inner ring of the supernova will be destroyed by $\sim$2025.  In the X-ray, \citet{Frank2016} report that the 0.5-2 keV light curve has remained constant since Day 9,500, and that the south-eastern side of the ER is fading, which is consistent with optical observations.  This suggests that the blast wave is indeed leaving the ER, and that the SNR is expanding into the surrounding region.  \citet{Potter2014} recently presented a model for radio emission which included the fact that the shock is expanding above and below the ER.

In this paper, we are interested in the most recent evolution of the SN 1987A radio remnant at 9 GHz, as part of a series of continuing monitoring \citep{Gaensler1997,Ng2008,Zanardo2013,Ng2013}.  The most recent observations of the SNR showed a potential break in the expansion of the remnant, which was perhaps due to a change in emission morphology \citep{Ng2013}.  Furthermore, a decreasing trend in the east-west surface emissivity of the remnant was observed, with predictions that the western lobe of the SNR would soon dominate in radio emission \citep{Ng2013,Potter2014}.  The eastern side of SN 1987A has always been brighter in earlier epochs, typically attributed to faster shocks in the east \citep{Gaensler1997, Zanardo2010}, which were possibly caused by an asymmetric explosion \citep{Zanardo2013}.  \citet{Ng2013} first noted the decrease in the ER asymmetry at Day 7,000, which was attributed to these faster shocks in the east encountering the ER, then slowing down and exiting the ER faster than shocks in the west.  We seek to identify whether these trends have continued, and also whether there is evidence of the shockwave leaving the ER as was reported at optical and X-ray wavelengths.

In this paper, we report on the updated radio observations of SN 1987A through Day 10,942 after the supernova explosion, using 9 GHz ATCA data from January 1992 to February 2017.  In Section \ref{sec:observations}, we discuss our observations and data reduction.  In Section 3, we show our resulting images, and discuss our analysis of the remnant including Fourier modeling techniques, and the subsequent results.  In Section \ref{sec:discussion} we discuss the physical implications of our results, and compare SN 1987A to the handful of other spatially resolved radio supernovae.

\section{Observations and Data Reduction}
\label{sec:observations}

Our radio observations are part of a continuing ATCA imaging project at 9 GHz of SN 1987A, which has been bright enough for imaging since Day 1,786 \citep{SS1993}.  These observations are taken using 6-km configurations only, with typically $\sim10$ hr of on-source time.  This publication covers all observations taken to February 2017 of the source, with observations prior to 2013 July 18 also discussed by \citet{Ng2013}.  Our new observations are summarized in Table \ref{tab:observations}.

Observations before 2009 in this monitoring campaign were made in two bands centered at 8.512 GHz and 8.896 GHz, respectively, with a usable bandwidth of 104 MHz each.  Since the Compact Array Broadband Backend \citep[CABB;][]{Wilson2011} upgrade, which enables observations on 2-GHz bandwidth on each band, we have restricted the analysis of 9-GHz monitoring data to two 104-MHz sub-bands that match the older observation settings.

The data were reduced using the MIRIAD software package \citep{Sault1995}, first using standard flagging and calibration techniques, and then implementing self-calibration \citep[see][]{Gaensler1997,Ng2008}.  We averaged the visibility data first in one-minute intervals, and formed intensity maps from the visibility data using a maximum entropy algorithm \citep{GD1978}.  Finally, we applied a super resolution technique by restoring the cleaned maps with a FWHM beam of 0.4" based on the method outlined by \citet{Briggs1994,Briggs1995}.

\startlongtable
\begin{deluxetable*}{lcclcc}
\tablecaption{Observations and parameters for the data sets in this study.  The line between Days 9,568 and 9,756 denotes new data unique to this work.}
\tabletypesize{\small}
\tablehead{
\colhead{Observing Date} & \colhead{Days since} & \colhead{Array} &
\colhead{Center Frequency\tablenotemark{a}} & \colhead{Time on} &
\colhead{Epoch Shown}  \\ \colhead{} & \colhead{Supernova} &
\colhead{Configuration} & \colhead{(MHz)} & \colhead{Source (hr)} &
\colhead{in Figure~\ref{fig:supernova-shots}\tablenotemark{b}}}
\startdata
1992 Jan 14& 1786 & 6B & 8640  & 12 & \nodata \\
1992 Mar 20& 1852 & 6A & 8640  & 10 & \nodata \\
1992 Oct 21& 2067 & 6C & 8640, 8900 & 13 & 1992.9 \\
1993 Jan 4 & 2142 & 6A & 8640, 8900 & 9 & 1992.9 \\
1993 Jan 5 & 2143 & 6A & 8640, 8900 & 6 & 1992.9 \\
1993 Jun 24& 2313 & 6C & 8640, 8900 & 8 & 1993.6 \\
1993 Jul 1 & 2320 & 6C & 8640, 8900 & 10 & 1993.6 \\
1993 Oct 15& 2426 & 6A & 8640, 9024 & 17 & 1993.6 \\
1994 Feb 16& 2550 & 6B & 8640, 9024 & 9 & 1994.4 \\
1994 Jun 27-28& 2681 & 6C & 8640, 9024 & 21 & 1994.4 \\
1994 Jul 1 & 2685 & 6A & 8640, 9024 & 10 & 1994.4 \\
1995 Jul 24& 3073 & 6C & 8640, 9024 & 12 & 1995.7 \\
1995 Aug 29& 3109 & 6D & 8896, 9152 & 7 & 1995.7 \\
1995 Nov 6 & 3178 & 6A & 8640, 9024 & 9 & 1995.7 \\
1996 Jul 21& 3436 & 6C & 8640, 9024 & 14 & 1996.7 \\
1996 Sep 8 & 3485 & 6B & 8640, 9024 & 13 & 1996.7 \\
1996 Oct 5 & 3512 & 6A & 8896, 9152 & 8  & 1996.7 \\
1997 Nov 11& 3914 & 6C & 8512, 8896 & 7  & 1998.0 \\
1998 Feb 18& 4013 & 6A & 8896, 9152 & 10 & 1998.0  \\
1998 Feb 21& 4016 & 6B & 8512, 9024 & 7  & 1998.0 \\
1998 Sep 13& 4220 & 6A & 8896, 9152 & 12 & 1998.9  \\
1998 Oct 31& 4268 & 6D & 8502, 9024 & 11 & 1998.9  \\
1999 Feb 12& 4372 & 6C & 8512, 8896 & 10 & 1999.7 \\
1999 Sep 5 & 4577 & 6D & 8768, 9152 & 11 & 1999.7 \\
1999 Sep 12& 4584 & 6A & 8512, 8896 & 14 & 1999.7 \\
2000 Sep 28& 4966 & 6A & 8512, 8896 & 10 & 2000.9 \\
2000 Nov 12& 5011 & 6C & 8512, 8896 & 11 & 2000.9 \\
2001 Nov 23& 5387 & 6D & 8768, 9152 & 8 & 2001.9 \\
2002 Nov 19& 5748 & 6A & 8512, 8896 & 8 & 2003.0 \\
2003 Jan 20& 5810 & 6B & 8512, 9024 & 9 & 2003.0 \\
2003 Aug 1 & 6003 & 6D & 8768, 9152 & 10 & 2003.6 \\
2003 Dec 5 & 6129 & 6A & 8512, 8896 & 9 & 2004.0 \\
2004 Jan 15& 6170 & 6A & 8512, 8896 & 9 & 2004.0 \\
2004 May 7 & 6283 & 6C & 8512, 8896 & 9 & 2004.4 \\
2005 Mar 25& 6605 & 6A & 8512, 8896 & 9 & 2005.2 \\
2005 Jun 21& 6693 & 6B & 8512, 8896 & 9 & 2005.5 \\
2006 Mar 28& 6973 & 6C & 8512, 8896 & 9 & 2006.2 \\
2006 Jul 18& 7085 & 6A & 8512, 8896 & 9 & 2006.5 \\
2006 Dec 8 & 7228 & 6B & 8512, 9024 & 8 & 2006.9 \\
2008 Jan 4 & 7620 & 6A & 8512, 9024 & 11 & 2008.0 \\ 
2008 Apr 23& 7730 & 6A & 8512, 8896 & 11 & 2008.3 \\
2008 Oct 11& 7901 & 6A & 8512, 8896 & 11 & 2008.8 \\
2009 Jun 6 & 8139 & 6A & 9000  & 11 &     2009.4 \\
2010 Jan 23& 8370 & 6A & 9000  & 11 &     2010.1 \\
2010 Apr 11& 8448 & 6A & 9000  & 11 &     2010.3 \\
2011 Jan 25& 8737 & 6A & 9000  & 11 &     2011.1 \\
2011 Apr 22& 8824 & 6A & 9000  & 11 &     2011.3 \\
2012 Jan 12& 9089 & 6A & 9000  & 11 &     2012.0 \\
2012 Jun 5 & 9233 & 6D & 9000  & 11 &     2012.4 \\
2012 Sep 1 & 9321 & 6A & 9000  & 10 &     2012.7 \\
2013 Mar 7 & 9509 & 6A & 90006  & 11 &     2013.2 \\
2013 May 5 & 9568 & 6C & 9000  & 11 &     2013.3 \\
\hline
2013 July 18 & 9642 & 6A & 8512, 8896 & 10 & 2013.5 \\
2013 Nov 09 & 9756 & 6A & 8512, 8896 & 9 & 2013.9 \\
2014 Apr 16 & 9915 & 6A & 8512, 8896 & 8 & 2014.3 \\
2016 Mar 03 & 10601 & 6B & 8512, 8896 & 8 & 2016.2 \\
2017 February 7 & 10942 & 6D & 8512, 8896 & 9 & 2017.2 \\
\enddata
\label{tab:observations}
\tablenotetext{a}{Since the CABB upgrade in mid-2009, data have been recorded
over a 2-GHz bandwidth. However, in this analysis we used two 104-MHz subbands
with center frequencies of 8.512\,GHz and 8.896\,GHz, for a consistency with
the bandwidth of pre-CABB data.}
\tablenotetext{b}{Some early datasets have been averaged together to generate the
corresponding images in Figure~\ref{fig:supernova-shots} for the listed epoch.}
\end{deluxetable*}

\section{Results} 
\label{sec:results}

\subsection{Images}
\label{sec:images}

\begin{figure*}
	\includegraphics[width=\textwidth,height=\textheight,,keepaspectratio]{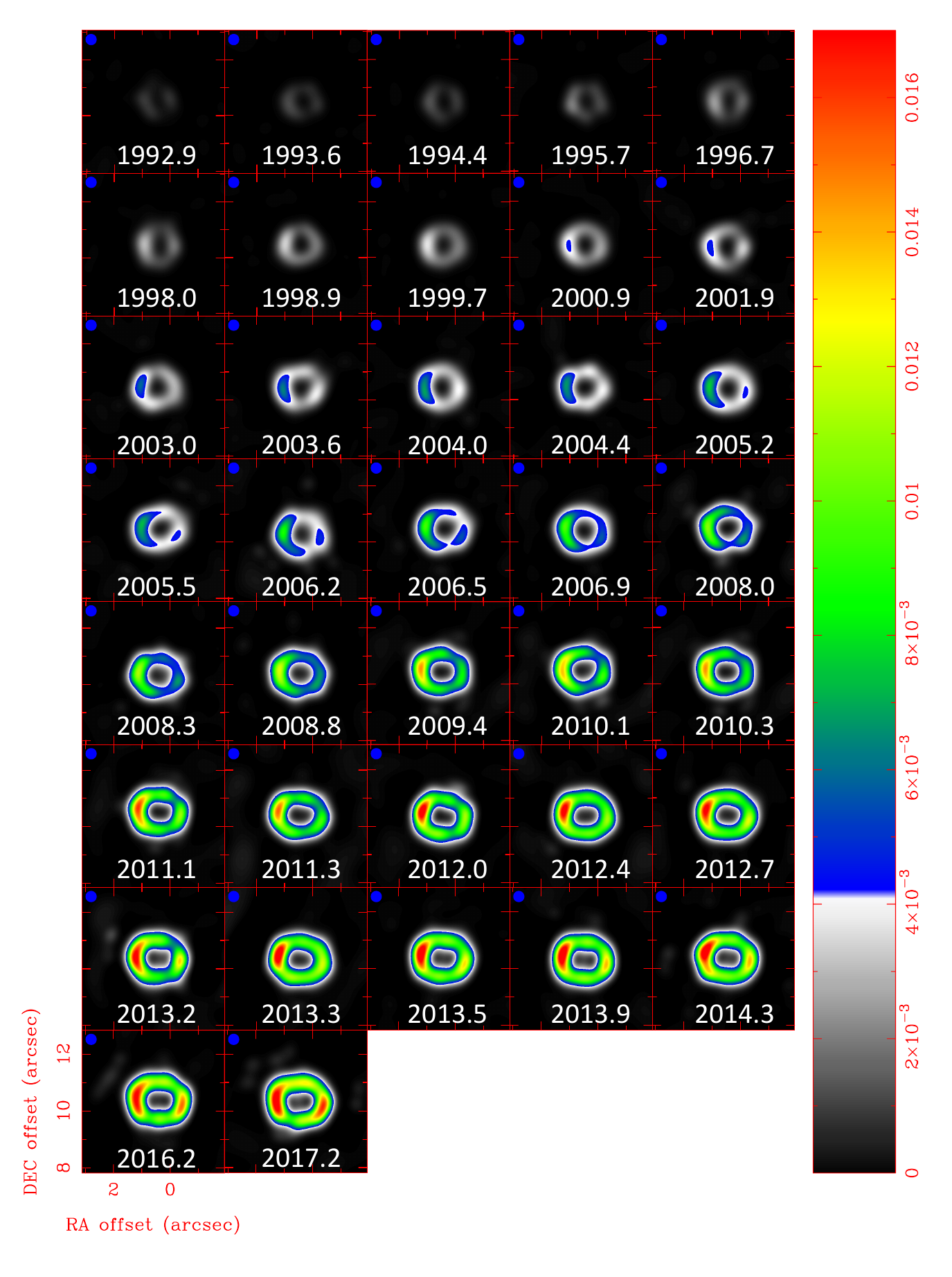}
    \caption{Super-resolved false color pixel images of SN 1987A at 9 GHz with ATCA data from 1992-2017.  North is up and east is left.  The scale on the right hand side is the intensity in Jy/beam, and we have provided the beam size for each image (blue circle, upper left hand corner) for reference.}
    \label{fig:supernova-shots}
\end{figure*}

The 9 GHz images of the remnant derived since 1992 are shown in Figure \ref{fig:supernova-shots}.  The remnant forms a circular shell, with brightness varying between different regions.  The left (east) side of the shell is consistently brighter than when it was first observed in 1992, and has steadily increased in brightness until the present day.  The right (west) side of the shell is increasing in brightness, but is less bright than the eastern side at all epochs.  Emission is also present but more limited in the top (north) and bottom (south) regions.  While emission is not uniform, we can see that all of the shell is increasing in brightness during our observations.  

We can also see a clear expansion in the shell from 1992 to present day.  In 1992, the radius of the emission region from the center of the SNR to the edge of emission in the eastern side can be measured directly from the image as $\sim$0.6", and the 2017 image shows expansion to $\sim$1.4".  A quantitative analysis of this expansion can be found in Section \ref{sec:expansion}.

\begin{figure*}
	\includegraphics[width=\textwidth]{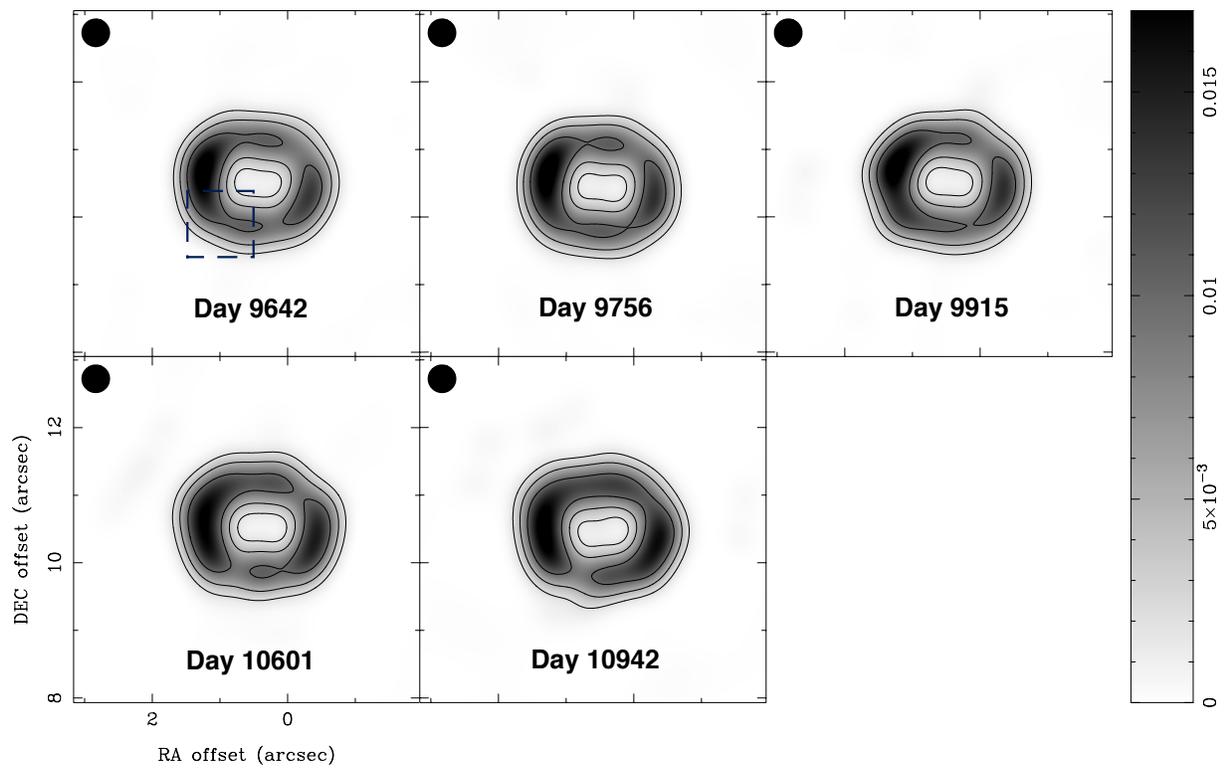}
    \caption{Contour plots of the last five images in Figure \ref{fig:supernova-shots}, by epoch of observation.  The contour levels correspond with 2, 5, 10, and 20 mJy/beam, respectively.  The circle in the upper left hand corner of each image is the beam size, and the dark blue box in the Day 9,642 image is the southeast corner region.}
    \label{fig:contours}
\end{figure*}

We have also included contour plots of the images that are new to this paper in Figure \ref{fig:contours}.  We can see that the circular structure of the remnant is present at all epochs, and that the remnant is getting brighter.  We also see that the south-eastern part of the remnant is the only region decreasing in brightness, which can be seen in the final two most recent observations.  Specifically, the average flux density in our first three images measured at a 135 degree angle, which is measured from north to east, (from Day 9,642, 9,756, and 9,915 respectively), is 10.0 $\pm$ 0.1 mJy, which fades to 9.0 $\pm$ 0.1 mJy by the observation on Day 10,061 and 8.2 $\pm$ 0.1 mJy on Day 10,942.  We acquired these values by measuring the flux density along the south-western region by drawing a box in the image plane using {\sc{kvis}}, which is marked in Figure \ref{fig:contours}.

\subsection{Fitting}
\label{sec:fitting}

The resolution of imaging data allows us to fit various models to the supernova shell structure and to obtain several parameters for its expansion.  For this, we used the Fourier modeling script outlined by \citet{Ng2008} and \citet{Ng2013}, which assesses the geometry of the remnant in the visibility ($uv$) plane, which is defined by a 2D Fourier transform of the sky brightness.  The fitting was done using a modified {\sc uvfit} task in MIRIAD, which in addition to the parameter values for the model also provides $\chi^{2}$ values in order to confirm the reliability of the fit.

We fit two geometries to the remnant: a 3D torus model and a 2D ring model.  Based on previous studies of 9-GHz observations \citep[e.g.,][]{Ng2008,Ng2013} and the complex remnant geometry (see simulations  by \citep{Potter2014}), we fit the data with two geometrical model: a 2D ring and a 3D torus.  The 2D models allow us to compare to the model used in X-ray studies \citep{Racusin2009,Helder2013,Frank2016}.  For the torus model, we fit for eight parameters: flux density, center position (in RA and Dec), radius, half-opening angle ($\theta$), thickness (as a fraction of the radius), asymmetry (as a percentage), and direction of a linear gradient in surface emission ($\phi$) \citep[see Section 3,][]{Ng2013}.  The parameters we obtained can be seen in Table 2 along with the confidence intervals from the $\chi^{2}$ distribution.

For the ring model, we fit the semi major axis ($R_{1}$) and semi-minor axis ($R_{2}$), as well as the flux density and center position\citep[see Section 3 of][]{Ng2013}.  The obtained parameters can be seen in Table 3 with the $\chi^{2}$ fit.

\clearpage
\startlongtable
\begin{deluxetable*}{lccccccr}
\tablecaption{Best-fit Parameters for the Torus Model with Statistical Uncertainties at 68\% Confidence Level.}
\tablewidth{\textwidth}
\label{tab:torus}
\tabletypesize{\small}
\tablehead{
\colhead{Day}&\colhead{Flux (mJy)}&\colhead{Radius (\arcsec)}
&\colhead{Half-opening}&\colhead{Thickness}&\colhead{Asymmetry}&\colhead{$\phi$ (\arcdeg)}&\colhead{$\chi^2_\nu$/dof\tablenotemark{a}}\\
&&&Angle, $\theta$ (\arcdeg)&\colhead{(\%)}&\colhead{(\%)}}
\startdata
1786& $4.2\pm0.2$ & $0.60\pm0.10$ & $84^{+6}_{-20} $ & $150\pm50$ & $70\pm30$ & $187\pm16$ &1.8/2107\\
1852& $4.0\pm0.3$ & $0.62\pm0.05$ & $80\pm10$ & $100\pm50$ & $100_{-60} $ & $180^{+10}_{-30}$ &3.7/1642\\
2067& $5.73\pm0.12$ & $0.62\pm0.05$ & $33\pm4$ & $175\pm25$ & $81\pm7$ & $121\pm6$ &4.3/3602\\
2142& $5.32\pm0.14$ & $0.65\pm0.02$ & $44\pm2$ & $172\pm10$ & $96\pm3$ & $114\pm7$ &17/2702\\
2143& $5.7\pm0.2$ & $0.64\pm0.01$ & $<12 $ &$<20$ & $40\pm8$ & $108\pm6$ &16/1392\\
2313& $6.73\pm0.11$ & $0.63\pm0.01$ & $34\pm7$ & $<20 $ & $40\pm5$ & $94\pm14$ &3.6/2902\\
2320& $7.04\pm0.13$ & $0.67\pm0.02$ & $37\pm7$ & $44^{+15}_{-20} $ & $38\pm5$ & $95\pm13$ &4.5/2962\\
2426& $6.65\pm0.10$ & $0.69\pm0.01$ & $55\pm4$ & $33^{+10}_{-16} $ & $42\pm5$ & $85\pm12$ &5.2/4372\\
2550& $6.63\pm0.15$ & $0.64\pm0.04$ & $26\pm5$ & $175\pm20$ & $80\pm6$ & $108\pm5$ &7.5/2992\\
2681& $8.41\pm0.08$ & $0.67\pm0.01$ & $48\pm3$ & $18^{+10}_{-17} $ & $33\pm4$ & $92\pm10$ &6.0/6142 \\
2685& $8.11\pm0.10$ & $0.66\pm0.01$ & $54\pm4$ & $<14 $ & $38\pm6$ & $113\pm12$ &5.7/3256\\
3073& $11.11\pm0.12$ & $0.67\pm0.01$ & $34\pm5$ & $46\pm14$ & $40\pm3$ & $93\pm8$ &5.6/2662\\
3109& $9.7\pm0.1$ & $0.64\pm0.02$ & $18^{+10}_{-18} $ & $<15 $ & $42\pm2$ & $88\pm7$ &17/1598 \\
3178& $11.71\pm0.09$ & $0.685\pm0.007$ & $45\pm2$ & $<10 $ & $39\pm2$ & $103\pm7$ &4/3442\\
3436& $15.17\pm0.09$ & $0.705\pm0.005$ & $47\pm2$ & $24\pm11$ & $42\pm2$ & $95\pm4$ &4.9/4337\\
3485& $15.42\pm0.08$ & $0.707\pm0.005$ & $51\pm2$ & $<10 $ & $43\pm2$ & $102\pm5$ &5.2/4702\\
3512& $15.43\pm0.12$ & $0.708\pm0.006$ & $53\pm2$ & $<18 $ & $42\pm3$ & $94\pm6$ &3.3/2632\\
3914& $17.57\pm0.14$ & $0.694\pm0.007$ & $42\pm3$ & $<10 $ & $38\pm3$ & $111\pm7$ &2.4/1272\\
4013& $19.09\pm0.10$ & $0.754\pm0.005$ & $46.4\pm1.4$ & $<5 $ & $45.1\pm1.5$ & $104\pm4$ &3.0/2830\\
4016& $18.72\pm0.10$ & $0.745\pm0.006$ & $51\pm2$ & $<5 $ & $46\pm2$ & $103\pm5$ &3.1/2512\\
4220& $20.20\pm0.09$ & $0.729\pm0.004$ & $43.4\pm1.3$ & $2^{+13}_{-2} $ & $37.6\pm1.3$ & $100\pm3$ &2.2/2955\\
4268& $21.78\pm0.13$ & $0.736\pm0.006$ & $40\pm2$ & $28\pm8$ & $38\pm2$ & $107\pm4$ &7.3/3862\\
4372& $22.94\pm0.10$ & $0.727\pm0.005$ & $37\pm2$ & $23\pm9$ & $37.4\pm1.5$ & $114\pm3$ &3.6/3532
\\
4577& $23.89\pm0.14$ & $0.757\pm0.007$ & $40\pm2$ & $21\pm6$ & $39\pm2$ & $103\pm4$ &7.0/3442\\
4584& $25.23\pm0.07$ & $0.747\pm0.003$ & $42.0\pm1.0$ & $<5 $ & $38.5\pm1.0$ & $109\pm2$ &2.9/4222\\
4966& $29.45\pm0.06$ & $0.764\pm0.002$ & $40.8\pm0.6$ & $<5 $ & $39.5\pm0.6$ & $108.3\pm1.3$ &1.3/50689\\
5011& $32.97\pm0.07$ & $0.775\pm0.002$ & $44.1\pm0.7$ & $1^{+5}_{-1} $ & $40.1\pm0.6$ & $104.9\pm1.4$ &1.4/50531\\
5387& $34.11\pm0.08$ & $0.790\pm0.003$ & $41.4\pm0.8$ & $<3 $ & $41.5\pm0.7$ & $107.8\pm1.5$ &1.6/39604\\
5748& $41.68\pm0.07$ & $0.811\pm0.002$ & $43.8\pm0.5$ & $<2 $ & $40.2\pm0.5$ & $103.1\pm1.2$ &1.3/38992\\
5810& $42.46\pm0.07$ & $0.815\pm0.002$ & $42.9\pm0.5$ & $1^{+4}_{-1} $ & $42.8\pm0.6$ & $117.4\pm1.0$ &1.6/46012\\
6003& $46.50\pm0.07$ & $0.815\pm0.002$ & $39.6\pm0.5$ & $<2 $ & $38.4\pm0.5$ & $101.2\pm1.1$ &1.6/44992\\
6129& $52.82\pm0.08$ & $0.833\pm0.002$ & $42.7\pm0.5$ & $<2 $ & $42.1\pm0.4$ & $107.7\pm1.1$ &1.3/46012\\
6170& $54.05\pm0.08$ & $0.831\pm0.002$ & $42.2\pm0.5$ & $<2 $ & $38.8\pm0.4$ & $107.8\pm1.1$ &1.5/43672\\
6283& $53.63\pm0.07$ & $0.829\pm0.001$ & $39.6\pm0.4$ & $<2 $ & $38.8\pm0.4$ & $107.7\pm0.9$ &1.5/44842\\
6605& $61.36\pm0.10$ & $0.843\pm0.002$ & $38.2\pm0.5$ & $<2 $ & $39.1\pm0.5$ & $109.0\pm1.1$ &2.9/42892\\
6693& $62.69\pm0.08$ & $0.858\pm0.001$ & $43.0\pm0.4$ & $<3 $ & $35.9\pm0.4$ & $101.5\pm1.0$ &1.5/38992\\
6973& $73.81\pm0.08$ & $0.880\pm0.001$ & $44.6\pm0.3$ & $<2 $ & $42.1\pm0.4$ & $117.4\pm0.7$ &1.5/40357\\
7085& $77.19\pm0.08$ & $0.872\pm0.001$ & $39.3\pm0.3$ & $<1 $ & $39.8\pm0.3$ & $111.8\pm0.6$ &1.5/29112\\
7228& $82.51\pm0.08$ & $0.874\pm0.001$ & $40.0\pm0.3$ & $<1$ & $39.4\pm0.3$ & $109.1\pm0.7$ &1.4/35677\\
7620& $93.61\pm0.09$ & $0.893\pm0.001$ & $42.7\pm0.3$ & $<4 $ & $38.9\pm0.3$ & $105.1\pm0.6$ &1.3/42892\\
7730& $98.98\pm0.08$ & $0.8905\pm0.0008$ & $36.2\pm0.2$ & $<1 $ & $36.3\pm0.2$ & $109.4\pm0.5$ &1.6/40942\\
7901& $107.73\pm0.08$ & $0.8916\pm0.0007$ & $35.8\pm0.2$ & $<2 $ & $35.9\pm0.2$ & $109.1\pm0.4$ &1.5/44452\\
8139& $121.59\pm0.08$ & $0.9095\pm0.0006$ & $37.6\pm0.1$ & $<4 $ & $32.0\pm0.1$ & $104.4\pm0.4$ &0.5/278820\\
8370& $128.07\pm0.08$ & $0.9142\pm0.0006$ & $36.6\pm0.2$ & $<1 $ & $33.4\pm0.2$ & $105.4\pm0.4$ &0.8/254302\\
8448& $132.59\pm0.07$ & $0.9109\pm0.0008$ & $33.5\pm0.2$ & $12\pm3$ & $28.7\pm0.1$ & $103.4\pm0.4$ &0.9/301177\\
8737& $142.25\pm0.07$ & $0.9185\pm0.0005$ & $32.8\pm0.1$ & $<1 $ & $29.8\pm0.1$ & $104.2\pm0.4$ &0.6/251707\\
8824& $136.58\pm0.08$ & $0.9169\pm0.0008$ & $28.4\pm0.2$ & $1^{+3}_{-1} $ & $26.9\pm0.1$ & $105.7\pm0.4$ &0.7/241327\\
9089& $155.92\pm0.06$ & $0.9251\pm0.0005$ & $30.5\pm0.1$ & $<1 $ & $25.8\pm0.1$ & $102.3\pm0.4$ &1.1/265987\\
9233& $161.00\pm0.06$ & $0.9290\pm0.0006$ & $28.8\pm0.1$ & $1^{+2}_{-1} $ & $25.8\pm0.1$ & $102.4\pm0.3$ &1.0/301177\\
9321& $165.62\pm0.05$ & $0.9304\pm0.0003$ & $29.1\pm0.1$ & $<1 $ & $23.5\pm0.1$ & $98.4\pm0.3$ &0.9/284272\\
9509& $168.68\pm0.05$ & $0.9467\pm0.0003$ & $31.7\pm0.1$ & $<1 $ & $26.7\pm0.1$ & $97.8\pm0.2$ &1.4/288550\\
9568& $176.01\pm0.04$ & $0.9378\pm0.0002$ & $28.6\pm0.1$ & $<1 $ & $18.4\pm0.1$ & $111.6\pm0.3$ &1.3/283341 \\
		9,642 & 175.81 $\pm$ 0.04 & 0.9464 $\pm$ 0.0002 & 29.68 $\pm$ 0.05 & <1 & 21.30 $\pm$ 0.04 & 99.6 $\pm$ .2 & 0.693/ 234782 \\
		9,756 & 179.87 $\pm$ 0.04 & 0.9508 $\pm$ 0.0002 & 30.84 $\pm$ 0.06 & <1 & 21.98 $\pm$ 0.05 & 103.5 $\pm$ 0.2 & 0.764/ 233190 \\
		9,915 & 185.96 $\pm$ 0.0005 & 0.9583 $\pm$ 0.0003 & 30.55 $\pm$ 0.06 & <2 & 20.57 $\pm$ 5 & 94.7 $\pm$ 0.3 & 0.7947/230290 \\
		10,601 & 199.45$\pm$ .06 & 0.9857 $\pm$ 0.0003& 32.48 $\pm$ 0.07 & <7 & 15.42 $\pm$ 0.5 & 88.6 $\pm$ 0.4 &0.578/118446 \\
		10,942 & 216.43 $\pm$ 0.06 & 0.9975 $\pm$ 0.0002 & 28.36 $\pm$ 0.05 & 9 $\pm$ 8 & 10.36 $\pm$ 6 & 87.5 $\pm$ 0.5 &  0.546/190019 \\
\enddata
\tablenotetext{a}{Before 2000, all 26 frequency channels in the data were
averaged into one band of effective bandwidth 208\,MHz to boost the signal;
between 2000 and 2009, 26 Hanning-smoothed channels, each of width 8\,MHz were
used in the fit; since mid-2009, after the installation of the Compact Array
Broadband Backend (CABB), 208 channels in the same frequency range were extracted,
each of width 1\,MHz. Since 2012, the ATCA sensitivity has improved by
$\sim$40\% as a result of the installation of new receivers.}
\end{deluxetable*}

\startlongtable
\begin{deluxetable*}{lccccccr}
\tablecaption{Best-fit Parameters for the Ring Model with Statistical
Uncertainties at 68\% Confidence Level}
\label{table:ring}
\centering
\tablewidth{\textwidth}
\tabletypesize{\small}
\tablehead{
\colhead{Day}&\colhead{Flux (mJy)}&\colhead{Semi-major}&
\colhead{Semi-minor}&\colhead{Asymmetry}&\colhead{$\phi$
(\arcdeg)}&\colhead{$\chi^2_\nu$/dof\tablenotemark{a}}\\
&&\colhead{Axis (\arcsec)}&\colhead{Axis (\arcsec)}&
\colhead{(\%)} &}
\startdata
1786& $3.70\pm0.12$ & $0.55\pm0.03$ & $0.50\pm0.03$ & $33\pm16$ & $141\pm26$ &1.8/2108\\
1852& $3.59\pm0.14$ & $0.53\pm0.04$ & $0.48\pm0.04$ & $25\pm17$ & $124\pm46$ &3.7/1643\\
2067& $5.17\pm0.09$ & $0.57\pm0.02$ & $0.49\pm0.01$ & $27\pm5$ & $105\pm16$ &4.4/3603\\
2142& $4.84\pm0.11$ & $0.53\pm0.02$ & $0.51\pm0.02$ & $17\pm8$ & $132\pm29$ &17/2703\\
2143& $5.62\pm0.14$ & $0.63\pm0.03$ & $0.44\pm0.02$ & $39\pm8$ & $94\pm24$ &16/1393\\
2313& $6.68\pm0.09$ & $0.57\pm0.02$ & $0.44\pm0.01$ & $34\pm5$ & $90\pm15$ &3.6/2903\\
2320& $6.87\pm0.10$ & $0.61\pm0.02$ & $0.47\pm0.01$ & $31\pm4$ & $92\pm13$ &4.5/2963\\
2426& $6.47\pm0.08$ & $0.56\pm0.01$ & $0.50\pm0.01$ & $28\pm4$ & $86\pm9$ &5.2/4373\\
2550& $6.12\pm0.12$ & $0.58\pm0.02$ & $0.54\pm0.02$ & $32\pm6$ & $117\pm12$ &7.5/2993\\
2681& $8.23\pm0.06$ & $0.58\pm0.01$ & $0.48\pm0.01$ & $26\pm3$ & $87\pm8$ &6.0/6143\\
2685& $7.94\pm0.08$ & $0.56\pm0.01$ & $0.49\pm0.01$ & $29\pm4$ & $109\pm10$ &5.7/3257\\
3073& $10.86\pm0.10$ & $0.62\pm0.01$ & $0.47\pm0.01$ & $33\pm2$ & $87\pm8$ &5.6/2663\\
3109& $9.73\pm0.23$ & $0.60\pm0.01$ & $0.50\pm0.01$ & $36\pm8$ & $92\pm32$ &17/1599\\
3178& $11.50\pm0.07$ & $0.601\pm0.007$ & $0.487\pm0.006$ & $32\pm2$ & $98\pm6$ &4.0/3443\\
3436& $14.76\pm0.07$ & $0.607\pm0.005$ & $0.495\pm0.004$ & $33\pm1$ & $94\pm4$ &5.0/4338\\
3485& $15.07\pm0.07$ & $0.598\pm0.006$ & $0.510\pm0.004$ & $33\pm2$ & $98\pm4$ &5.2/4703\\
3512& $15.06\pm0.10$ & $0.592\pm0.007$ & $0.508\pm0.005$ & $32\pm2$ & $91\pm5$ &3.4/2633\\
3914& $17.31\pm0.12$ & $0.624\pm0.006$ & $0.491\pm0.007$ & $33\pm3$ & $107\pm7$ &2.4/1273\\
4013& $18.53\pm0.09$ & $0.646\pm0.005$ & $0.527\pm0.004$ & $35.5\pm1.2$ & $99\pm3$ &3.0/2831\\
4016& $18.36\pm0.09$ & $0.632\pm0.005$ & $0.545\pm0.006$ & $35.8\pm1.3$ & $98\pm4$ &3.1/2513\\
4220& $19.76\pm0.07$ & $0.643\pm0.004$ & $0.509\pm0.003$ & $31.4\pm1.0$ & $96\pm3$ &2.2/2956\\
4268& $21.19\pm0.11$ & $0.656\pm0.006$ & $0.506\pm0.005$ & $31.1\pm1.5$ & $104\pm4$ &7.4/3863\\
4372& $22.43\pm0.08$ & $0.663\pm0.004$ & $0.500\pm0.004$ & $32.1\pm1.2$ & $111\pm3$ &3.6/3533\\
4577& $23.26\pm0.12$ & $0.674\pm0.006$ & $0.515\pm0.005$ & $32.7\pm1.4$ & $98\pm4$ &7.1/3443\\
4584& $24.69\pm0.07$ & $0.663\pm0.003$ & $0.521\pm0.003$ & $31.6\pm0.8$ & $103\pm2$ &2.9/4223\\
4966& $28.75\pm0.05$ & $0.681\pm0.002$ & $0.527\pm0.002$ & $32.9\pm0.5$ & $102\pm1$ &1.3/50690\\
5011& $32.03\pm0.06$ & $0.677\pm0.002$ & $0.531\pm0.002$ & $32.7\pm0.5$ & $100\pm1$ &1.4/50532\\
5387& $33.19\pm0.07$ & $0.694\pm0.002$ & $0.547\pm0.002$ & $34.0\pm0.5$ & $102\pm1$ &1.6/39605\\
5748& $40.56\pm0.06$ & $0.710\pm0.002$ & $0.564\pm0.001$ & $33.0\pm0.4$ & $97\pm1$ &1.3/38993\\
5810& $41.24\pm0.06$ & $0.723\pm0.002$ & $0.568\pm0.002$ & $34.8\pm0.4$ & $110\pm1$ &1.6/46013\\
6003& $45.35\pm0.06$ & $0.723\pm0.002$ & $0.561\pm0.002$ & $32.2\pm0.4$ & $94\pm1$ &1.6/44993\\
6129& $51.43\pm0.07$ & $0.733\pm0.002$ & $0.579\pm0.002$ & $34.7\pm0.3$ & $100\pm1$ &1.3/46013\\
6170& $52.63\pm0.07$ & $0.734\pm0.001$ & $0.577\pm0.002$ & $32.1\pm0.3$ & $102\pm1$ &1.5/43673\\
6283& $52.18\pm0.06$ & $0.743\pm0.001$ & $0.569\pm0.001$ & $32.8\pm0.3$ & $101\pm1$ &1.5/44843\\
6605& $59.86\pm0.09$ & $0.758\pm0.002$ & $0.586\pm0.002$ & $33.0\pm0.4$ & $101\pm1$ &2.9/42893\\
6693& $61.21\pm0.07$ & $0.7622\pm0.0013$ & $0.5979\pm0.0013$ & $30.1\pm0.3$ & $95\pm1$ &1.5/38993\\
6973& $71.33\pm0.07$ & $0.7743\pm0.0011$ & $0.6118\pm0.0014$ & $33.7\pm0.3$ & $110\pm1$ &1.5/40358\\
7085& $74.60\pm0.07$ & $0.7695\pm0.0012$ & $0.5939\pm0.0011$ & $32.6\pm0.2$ & $103\pm1$ &1.6/29113\\
7228& $80.53\pm0.07$ & $0.7825\pm0.0009$ & $0.6171\pm0.0010$ & $33.0\pm0.2$ & $101\pm1$ &1.4/35678\\
7620& $90.47\pm0.08$ & $0.7830\pm0.0011$ & $0.6208\pm0.0009$ & $31.8\pm0.2$ & $97\pm1$ &1.4/42893\\
7730& $96.36\pm0.07$ & $0.8058\pm0.0008$ & $0.6179\pm0.0008$ & $30.9\pm0.2$ & $102\pm1$ &1.6/40943\\
7901& $104.94\pm0.06$ & $0.8075\pm0.0007$ & $0.6220\pm0.0007$ & $30.8\pm0.1$ & $102\pm0$ &1.5/44453\\
8139& $117.73\pm0.07$ & $0.8241\pm0.0006$ & $0.6263\pm0.0006$ & $27.7\pm0.1$ & $97\pm0$ &0.5/278821 \\
8370& $124.86\pm0.07$ & $0.8243\pm0.0006$ & $0.6428\pm0.0007$ & $28.6\pm0.1$ & $98\pm0$ &0.8/254303 \\
8448& $129.50\pm0.06$ & $0.8406\pm0.0005$ & $0.6282\pm0.0006$ & $25.6\pm0.1$ & $94\pm0$ &0.9/301178 \\
8737& $139.17\pm0.06$ & $0.8433\pm0.0005$ & $0.6430\pm0.0005$ & $26.2\pm0.1$ & $96\pm0$ &0.7/251708 \\
8824& $133.95\pm0.07$ & $0.8582\pm0.0006$ & $0.6373\pm0.0007$ & $24.1\pm0.1$ & $96\pm0$ &0.7/241328 \\
9089& $153.09\pm0.05$ & $0.8593\pm0.0004$ & $0.6438\pm0.0005$ & $23.1\pm0.1$ & $92\pm0$ &1.1/265988 \\
9233& $157.94\pm0.05$ & $0.8698\pm0.0004$ & $0.6401\pm0.0005$ & $23.4\pm0.1$ & $93\pm0$ &1.1/301178 \\
9321& $162.98\pm0.05$ & $0.8703\pm0.0003$ & $0.6527\pm0.0004$ & $21.5\pm0.1$ & $88\pm0$ &0.9/284273 \\
9509& $165.10\pm0.04$ & $0.8761\pm0.0003$ & $0.6583\pm0.0003$ & $24.0\pm0.1$ & $88\pm0$ &1.4/288551 \\
9568& $173.03\pm0.04$ & $0.8818\pm0.0002$ & $0.6574\pm0.0003$ & $16.2\pm0.1$ & $100\pm0$ &1.4/283342 \\
9642 & 171.54$\pm$ .04 & 0.8781 $\pm$ 0.0002 & 0.6580 $\pm$ 0.0002 & 19.28 $\pm$ 0.04 & 81.7 $\pm$ 0.2 &0.762/258312 \\
9756 & 176.54$\pm$ .04 & 0.8810 $\pm$ 0.0002 & 0.669 $\pm$ 0.0003 & 19.53 $\pm$ 0.04 & 93.1 $\pm$ 0.1 &0.837/4255481 \\
9915 & 181.69 $\pm$ .05 & 0.8916 $\pm$ 0.0003 & 0.663 $\pm$ 0.0003 & 18.87 $\pm$ 0.05 & 89.1 $\pm$ 0.3 &0.655/152784 \\
10601 & 195.68$\pm$ .06 & 0.9182 $\pm$ 0.0003 & 0.691 $\pm$ 0.0003 & 14.61 $\pm$ 0.06 & 74.1 $\pm$ 0.4 &0.651/133454 \\
10942 & 209.78 $\pm$ .06 & 0.9348 $\pm$ 0.0003 & 0.634 $\pm$ 0.0003 & 10.64 $\pm$ 0.07 & 64.1 $\pm$ 0.4 & 0.653/227191 \\
\enddata 
\tablenotetext{a}{Before 2000, all 26 frequency channels in the data were
averaged into one band of effective bandwidth 208\,MHz to boost the signal;
between 2000 and 2009, 26 Hanning-smoothed channels, each of width 8\,MHz were
used in the fit; since mid-2009, after the installation of the Compact Array
Broadband Backend (CABB), 208 channels in the same frequency range were extracted,
each of width 1\,MHz. Since 2012, the ATCA sensitivity has improved by
$\sim$40\% as a result of the installation of new receivers.}
\end{deluxetable*}

\subsection{Expansion}
\label{sec:expansion}

\begin{figure}
	\includegraphics[width=\columnwidth]{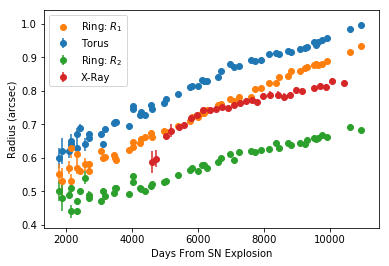}
    \caption{The radius of SN 1987A as a function of time, as measured using the torus model (blue), and the semi-major ($R_{1}$) and semi-minor ($R_{2}$) axes for the ring model (yellow and green, respectively).  The radius as measured in the X-ray by Chandra (red) is provided for reference from \citet{Frank2016}.}
    \label{fig:radius-all}
\end{figure}

The obtained radii for the torus model, and for the ring model, can be seen in Figure \ref{fig:radius-all}.  We also included the X-ray radius reported by \citet{Frank2016} for reference, who fit their data with a ring model using a single radius.  All three of our models show a clear increase in the radius over the entire time period.

\begin{figure}[h]
  \begin{center}
    \subfigure[Torus Radii]{\label{fig:edge-a}\includegraphics[scale=0.45]{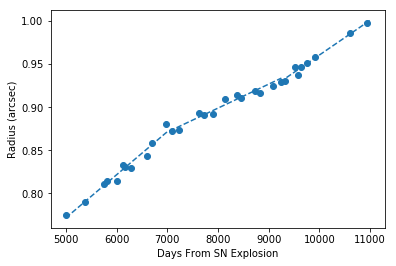}}
    \subfigure[Semi-Major Axis in Ring fit ($R_{1}$)]{\label{fig:edge-b}\includegraphics[scale=0.45]{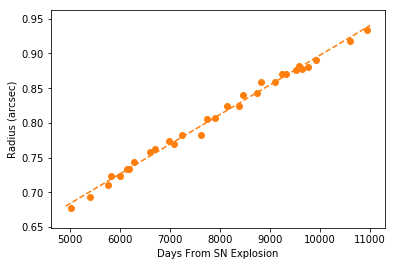}} \\
    \subfigure[Semi-Minor Axis in Ring ($R_{2}$)]{\label{fig:edge-c}\includegraphics[scale=0.45]{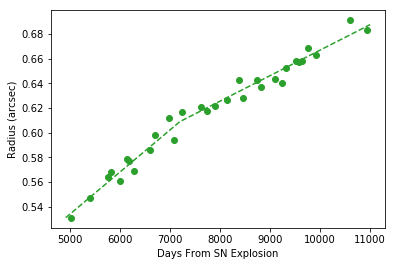}}
  \end{center}
  \caption{A subset of data shown in Figure \ref{fig:radius-all} to highlight the difference in fit between the torus radii, $R_{1}$, and $R_{2}$.  Here, we show the multi-linear fit for the torus radius from Day 7,500, and the linear fit to $R_{1}$ and $R_{2}$ from Day 7,000.}
  \label{fig:radii-fits}
\end{figure}

\begin{table*}
	\centering
	\begin{tabular}{llc}
		\hline
		Model & Type & BIC \\
		\hline
		Torus & Piecewise & -368.7  \\
		Torus & Linear &  -325.1 \\
		Ring ($R_{1}$) & Piecewise & -373.7  \\
		Ring ($R_{1}$) & Linear  & -378.8 \\
		Ring ($R_{2}$) & Piecewise (semi-minor axis) & -365.6   \\
		Ring ($R_{2}$) & Linear (semi-minor axis) &  -338.7 \\
		\hline
	\end{tabular}
	\caption{Bayes factors calculated via the Bayesian Information Criterion (BIC) for six linear and multi-linear (piecewise) models of the radio remnant that rely on three radius geometry parameters (torus, semi-major radius $R_1$, semi-minor radius $R_2$). All models fit data from Day 7200 to Day 10942.  Smaller BIC values indicate more accurate fits.}
	\label{tab:stats-results}
\end{table*}

\citet{Ng2013} analyzed the SNR expansion data from Day 4,000, and reported a break in the expansion rate for the torus model at Day 7,000 $\pm$ 200 to a lower speed (with a similar break occurring at the same time in $R_{2}$), and a linear trend in $R_{1}$.  Thus, when examining our most current data, first we examined data from Day 4,000 onwards in our analysis, as we were most interested in recent changes in the expansion of the emission.  For each of our models, we considered both a linear fit from Day 4,000 to present, and a break in expansion where the initial rate of linear expansion of the emission region changes to a second rate of expansion.  In the latter case, we fit a piecewise function consisting of two different linear slopes, and the transition day where the slope changed was fit as a free parameter.  This data fit is consistent with the earlier analysis of \citet{Ng2013}, and should emphasize that we do not believe these times don't necessarily refer to physical events, but rather this method was used in order to identify roughly where changes in the expansion rate occurred.  For each fit, we then calculated the Bayesian Information Criterion \citep[BIC;][]{Hogg2010} so we could compare the two models for each data set (Table \ref{tab:stats-results}), where the lower value indicates a more accurate fit.  We chose this over the $\chi^{2}$ fit because the broken linear fit has an extra parameter when compared to a simple linear fit, and the BIC takes into account potential overfitting with this method.  As in \citep{Kass1995}, given $\Delta B_ij = | B_i - B_j |$, where $B_i$ and $B_j$ are the BIC values associated with two statistical models, if $ 3 < \Delta B_ij < 10 $, the smaller BIC value provides \textit{substantial} evidence that the associate model is more accurate. For $\Delta B_ij>10$,  the smaller BIC value provides \textit{strong} evidence that the associate model is more accurate. Therefore, the BIC values listed in Table \ref{tab:stats-results} indicate that the torus model is the most accurate representation of the data after Day 7000, since $\Delta B_ij = | B_4 - B_2 | = 43.6.$

\begin{table*}
	\centering
	\begin{tabular}{lccccr}
		\hline
		Model & & Transition Date (Days From SN) & & Velocity (km/s)  \\
		& Transition 1 & Transition 2 & $\mathrm{v_{1}}$ &  $\mathrm{v_{2}}$ & $\mathrm{v_{3}}$ \\
		\hline
		Torus & 7,000 $\pm$ 200 & 9,300 $\pm$200 & 4,600$\pm$200 & 2,300$\pm$200 & 3,610$\pm$240 \\
		Ring: semi-major axis ($R_{1}$) & ---  & --- & 3,800 $\pm$460 &---& --- \\
		Ring: semi-minor axis ($R_{2}$) & 7,000 $\pm$300  & --- & 3,300$\pm$200 & 2,120$\pm$40 & --- \\
		\hline
	\end{tabular}

	\caption{Expansion velocities and errors of the radio remnant from Day 4,000 over time. Values for $v_{1}$ were obtained from \citet{Ng2013}, $v_{2}$ corresponds with the velocity after the first transition date, and $v_{3}$ corresponds with the velocity after the second transition date.  A dash corresponds with when no transition to a different velocity was observed.}
	\label{tab:expansion}
\end{table*}

From this, our results for the expansion velocities can be seen in Table \ref{tab:expansion}, where we assumed a distance to the supernova of 51.4 kpc to obtain expansion velocities in km/s.    We found that during this entire period, $R_{1}$ has continued expanding linearly of 3,800 $\pm$460 km/s, which is in agreement with the reported value in \citet{Ng2013} of 3,890$\pm$50 km/s.  Our test also confirmed the break previously observed at Day 7,000 in the cases of both the torus model and $R_{2}$.  Based on this, we further examined the data from Day 7,000 to the present with another BIC test comparing a broken linear versus linear model.  In this case, linear expansion was the best model for $R_{2}$ from Day 7,000, with a velocity of 2,120$\pm$40 km/s, which is greater than the previously reported value for Day 7,000-9,568 from \citet{Ng2013} of 1,750$\pm$300 km/s).  For the torus model, we identified a second break in the expansion velocity that best fits our data from Day 7,000 to the present period.  This occurred at Day 9,300 $\pm$ 200, where the expansion rate of the torus model changed from 2,300$\pm$200 km/s to 3,610$\pm$240 km/s.  For reference, the best fits to our data are in Figure \ref{fig:radii-fits}.

\subsection{Radio Light Curve}
\label{sec:light-curve}

\begin{figure}
	\includegraphics[width=\columnwidth]{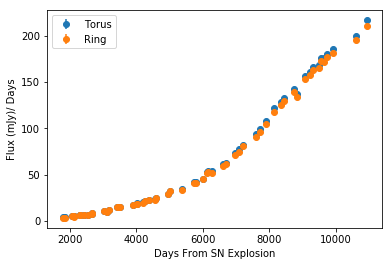}
    \caption{The measured flux densities for the SN 1987A emission region for both the torus and ring models.  Note that the errors are included in this plot, but are too small to be visible.}
    \label{fig:flux-scatter}
\end{figure}

The integrated flux densities obtained via  the torus and ring models are compared in Figure \ref{fig:flux-scatter}.  We chose to measure and list the flux of the model, not the source itself, because this value is tied to the overall geometry versus the individual fluctuations and our goal was to understand how the models can change.  For consistency, we checked the flux of the source in kvis for the first and last natural images in our data set.  For the Day 1,786 date, we found a flux density of 4.3 $\pm$ .1 mJy, which can be compared to 4.2 $\pm$ .2 mJy for the torus model and 3.70 $\pm$ .12 in the ring model.  On Day 10,942, we measure 212 $\pm$ 3 mJy with this method, compared to 216.43 $\pm$ .06 mJy for the torus model and 209.78 $\pm$ .06 mJy for the ring model.  This shows our using this method to measure total flux is consistent with measuring it for the entire source region, and can even be more accurate.

We find that the ring and torus flux density were in agreement, except the ring model is lower than the flux density measurement when compared to the torus measurement in later epochs.  The discrepancy between the two values, which begins at Day $\sim$6,000, has increased over time, to a $\sim$7 mJy difference in 2017, or 3\% of the total.  Measuring the flux density using another method such as the peak flux value using a conventional software package like {\sc{kvis}} could not distinguish between the two models, as the uncertainty from this method is larger than the difference in flux density between the two models.  Instead, we examined the maps of the residual visibilities (i.e., our data minus the model) for both the torus and ring models.  We found the torus model accounted for the emission more accurately than the ring model by showing less residual flux.  This leads us to conclude that the flux emission from the torus model is more accurate than the ring model, although we include values for the ring model here for completeness.

\begin{figure}
	\includegraphics[width=\columnwidth]{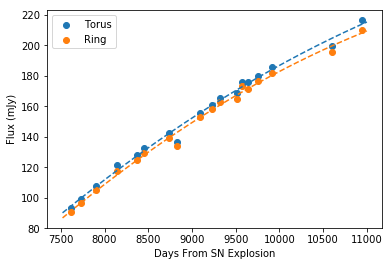}
    \caption{The measured flux densities for the SN 1987A emission region for both the torus and ring models from Day 7,500, along with the exponential fits to the data following Equation \ref{eq:lightcurve} and Table \ref{tab:exp-fit}.}
    \label{fig:flux-fit}
\end{figure}

\citet{Ng2013} reported that the fluxes obtained were lower than expected from the exponential fit given by \citet{Zanardo2010} from Day $\sim$7,500.  After this time, our data confirm a continuation of this trend.  As such, we considered data from Day 7,500 onward to derive a new modified power law fit.  The radio light curve of SN 1987A can be parameterized by

\begin{equation}
\mathrm{S(mJy) = K} \bigg(\frac{\nu}{5\:\mathrm{ GHz}}\bigg)^{\alpha}\bigg(\frac{t- t_{0}}{1\:\mathrm{ day}}\bigg)^{\beta} \,,
\label{eq:lightcurve}
\end{equation}

\begin{table*}
	\centering
	\begin{tabular}{lcccr}
		\hline
		Model & K & $t_{0}$ (days) & $\beta$ \\
		\hline
		Torus & 1.5 $\pm$ 0.1 & 6,540 $\pm$ 80 & 0.59 $\pm$ 0.02 \\
		Ring & 2.0 $\pm$ 0.12& 6,680 $\pm$  80& 0.55 $\pm$ 0.02 \\
		\hline
	\end{tabular}
	\caption{Parameters for the exponential fits seen in Figure \ref{fig:flux-fit}.}
	\label{tab:exp-fit}
\end{table*}

where $K$ is the parameter fit due to the flux density, $\nu$ is the frequency of observation, $\alpha$ is the spectral index, and $\beta$ is the modified power law fit for the light curve \citep{Weiler2002}.  Using $\nu=9$ GHz, $\alpha=-0.74$ \citep{Zanardo2013} , and the curve fitting package in \citep[{\sc{SciPy}};][]{SciPy}, we obtained the parameter values seen in Table \ref{tab:exp-fit}.  The data generated with this model along with the relevant fits can be seen in Figure \ref{fig:flux-fit} for both models.

\begin{figure}
	\includegraphics[width=\columnwidth]{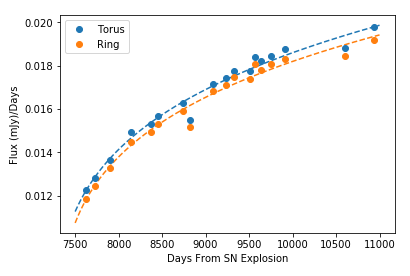}
    \caption{The same plot as Figure \ref{fig:flux-fit}, but with the fluxes divided by the day number.  This is done to highlight the fact that the rate of brightening in both models has been growing more slowly with time.}
    \label{fig:flux-day-fit}
\end{figure}

We also did the same fitting routine by dividing the flux values for each model by the number of days since the supernova explosion (Figure \ref{fig:flux-day-fit}), in order to see whether there are changes in the increase in flux density itself.  Here, we can see that from Day 7,500, the rate of brightness increase has been growing more slowly over time.

\subsection{Morphological Behavior}
\label{sec:phi}

\begin{figure}
	\includegraphics[width=\columnwidth]{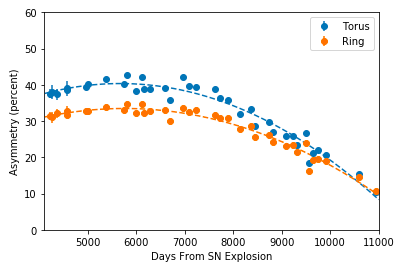}
    \caption{The east-west asymmetry in the SN 1987A emission region for both the torus and ring models, modeled as a linear gradient in emissivity across the the equatorial plane.  The parameters for the fits can be found in Table \ref{tab:asym-fit}}.
    \label{fig:asymmetry}
\end{figure}

\begin{table*}
	\centering
	\begin{tabular}{lcccr}
		\hline
		Model & a & $t_{0}$ (days) & $\beta$ \\
		\hline
		Torus & 4.0 $\pm$ 0.1 & 6,540 $\pm$ 80 & -1.129$\times 10^{-6}\pm 9\times10^{-9}$  \\
		Ring & 4.6 $\pm$ 0.2& 6,680 $\pm$  80& -8.71$\times 10^{-7}\pm 9\times10^{-9}$ \\
		\hline
	\end{tabular}
	\caption{Parameters for the asymmetry fits seen in Figure \ref{fig:asymmetry}.}
	\label{tab:asym-fit}
\end{table*}

Although the remnant of SN 1987A is increasing in brightness on a power law scale, this is not evenly distributed throughout the entire ring.  Surface brightness asymmetry has been apparent in the remnant since the earliest radio observations, with the eastern (left) lobe the brightest area.  We confirm that the eastern lobe is still increasing in flux, although emission from the southeastern quadrant is fading (Figure \ref{fig:contours}).

On the western side of the source, emission is increasing at a faster rate than previous years when compared to the eastern side.  This is particularly visible in the east-west asymmetry of surface emissivity, defined as the slope of the gradient of flux in the image plane (north to east), as seen in Figure \ref{fig:asymmetry}.  We note this decrease is greater in the ring model than in the torus model.  We provide a fit for the change in asymmetry gradient over time, the parameters for which can be seen in Table \ref{tab:asym-fit}.

\begin{figure}
	\includegraphics[width=\columnwidth]{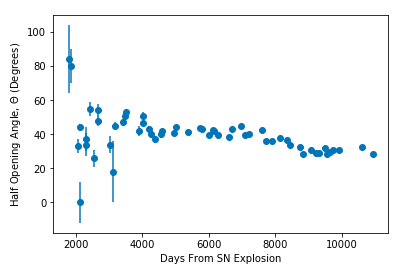}
    \caption{The measured half-opening angle ($\theta$) in the the SN 1987A remnant for the torus model.}
    \label{fig:theta}
\end{figure}

The torus model also provides us with the half-opening angle ($\theta$).  A plot of these data can be seen in Figure \ref{fig:theta}.  The orientation of the half opening angle has become more stable after Day 4,000, and between Day 4,000 and 7,700 has a mean value $\theta = 41.4 ^\circ$.  From this point, $\theta$ begins to rapidly decrease, until stabilizing again at Day 8,800 with a new mean value of $\theta = 29.9 ^\circ$.

\begin{figure}
	\includegraphics[width=\columnwidth]{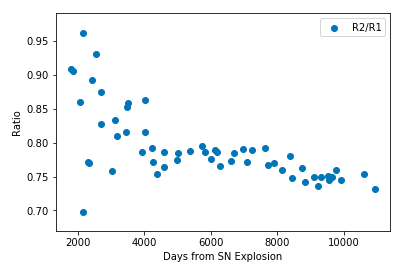}
    \caption{The measured ratio for $R_{1}$/$R_{2}$ for the ring model.  We will note that in this plot the errors are included, but are too small to be visible.}
    \label{fig:R1byR2}
\end{figure}

In the case of the ring model, \citet{Ng2013} noted that the ratio of $R_{2}$/$R_{1}$ closely follows the trends seen in $\theta$.  Results for this ratio over time can be seen in Figure \ref{fig:R1byR2}.  We find that during the Day 4,000- 7,700 period, the $R_{2}$/$R_{1}$ ratio is 0.78 $\pm$ 0.01, and after Day 8,800 (when $\theta$ stabilizes), the ratio is 0.74 $\pm$ 0.02.  As such, our findings are consistent with the results from \citet{Ng2013}, and have continued to Day 11,000.

\section{Discussion}
\label{sec:discussion}

The current radio emission from SN 1987A at 9 GHz is dominated by the interaction of the supernova shockwave with the ER, which is a smooth ring interspersed with denser clumps, plus an unknown contribution of polar emission from outside the dense ring.  We are currently witnessing the forward shock from the supernova explosion leaving the ER as the reverse shock is driven back into this ring.  We first discuss the picture of the SN 1987A CSM during the period when the supernova blast wave pushed through the CSM (Day 4,000-9,300), both in the morphological and physical contexts.

Our breakdown of the Day 4,000-9,300 period in Section \ref{sec:discussion} is as follows.  First, in Section \ref{sec:rateofexp} we discuss the expansion index of both the ring and torus models, which indicates that the increased expansion rate during this period is caused by shockwaves at high latitudes in the SNR.  In Section \ref{sec:exp-half-opening}, we discuss the coupling between $\theta$ and the expansion of the remnant.  In Section \ref{sec:compare-to-x-ray}, we compare our radio observations to X-ray data, and discuss how the difference in size between the two is likely physical.  In Section \ref{sec:Asymmetry}, we discuss the delayed change in the asymmetry between the eastern and western lobes of the SNR as predicted from theory and seen at other wavelengths.  And in Section \ref{sec:compare-to-x-ray}, we discuss our increasing flux brightness in the context of potential high latitude emission, and a potential change in emission region within the ER.

After this, in Section \ref{sec:9300} we will discuss the observed changes in SN 1987A from Day 9,300 onwards, which we interpret as the time at which the shockwave left the ER.  In Section \ref{sec:future}, we will discuss our future predictions for the SNR, including factors such as high latitude emission and the unknown composition of the region beyond the ER, and make future predictions for the change in asymmetry in emission.  Finally, in Section \ref{sec:compare}, we compare SN 1987A to other resolved radio supernovae, and discuss in this context how SN 1987A can be used to contrast features such as hotspots and expansion rates in various SNR.

\subsection{Day 4,000- Day 9,300}
\label{sec:4000-9300}

\subsubsection{Rate of Expansion}
\label{sec:rateofexp}

In Section \ref{sec:expansion} and Figure \ref{fig:radius-all}, we presented our results from the expansion of both the torus and ring models of SN 1987A.  When comparing our two models for the SNR expansion, there is a discrepancy in the values for the radius measurement between the torus model radius and $R_{1}$, where the torus values are consistently larger than $R_{1}$ (and $R_{2}$).  This discrepancy was first noted by \citet{Ng2013}, who described the cause of this phenomenon in more detail and presented two explanations for it.  The first is that the torus model, unlike the ring model, is dependent on the half-opening angle, $\theta$ (see Section \ref{sec:exp-half-opening}).  This then creates a projection effect, as the torus model approximates a ring when $\theta=0^{\circ}$, but is a shell as $\theta$ increases to $90^{\circ}$, and the shell model can have its emission peak inside the shell's radius.  Another proposed explanation is that because the shockwave travels faster in the lower density regions above the ER \citep{Blondin1996}, any high-latitude material present would manifest in a higher radius measurement for the 3D-sensitive torus model.

\begin{figure}
	\includegraphics[width=\columnwidth]{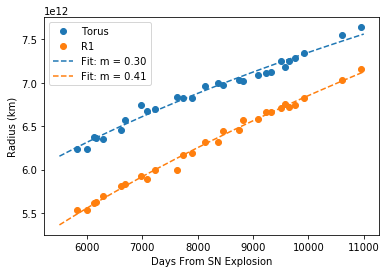}
    \caption{Radius as a function of time during the period when the shockwave was interacting with the ER.  We fit the data to $R \propto t^{m}$, which resulted in a value for m$=0.30 \pm 0.01$ for the torus model m$=0.41 \pm 0.01$ for $R_{1}$.}
    \label{fig:exp-fit}
\end{figure}

In the case of spherical symmetry in the SNR and a power law density distribution, one can model the expansion of the supernova remnant as $R \propto t^{m}$ \citep{Chevalier1982,Chevalier1982B}.  This is because we expect the forward and reverse shocks to drive our observed expansion, and we can expect these shocks to slow down over time as the ejecta transfer some amount of their kinetic energy into the swept-up material.  We can assume the shockwave entered the ER by Day 5,600 \citep{Helder2013}, and from this point to the transition observed at Day 9,300 (see Section \ref{sec:9300}), we find m$=0.30 \pm$ 0.01 for the torus data, and 0.41 $\pm$ 0.01 for $R_{1}$.  

\subsubsection{Expansion and Half-Opening Angle}
\label{sec:exp-half-opening}

\begin{figure}
	\includegraphics[width=\columnwidth]{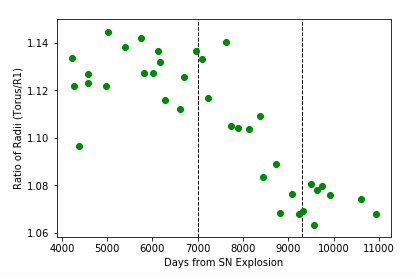}
    \caption{The ratio of the torus radii to $R_{1}$ from Day 4,000 to present.  The black dashed lines are at the two breaks in velocity detected in the torus radius at Day 7,000 and Day 9,300, respectively.}
    \label{fig:torusandR1}
\end{figure}

When considering the expansion rate of the radio remnant, we already explained in Section \ref{sec:rateofexp} that the torus and ring models give us different values for the radius.  However, the ratio between the torus radius and $R_{1}$ is not constant, due to the changing velocity of the torus radius (Figure \ref{fig:radius-all}).  Figure \ref{fig:torusandR1} shows the changing ratio of the torus radius divided by $R_{1}$ from Day 4,000.  For reference, we have indicated the deceleration at Day 7,000 and re-acceleration at Day 9,300 (as measured from the torus model) with black dashed lines (See Table \ref{tab:expansion}).  

Before Day 7,000, we find that the ratio is 1.13 $\pm$ 0.01.  This corresponds with a period during which \citet{Ng2013} reports that the torus model velocity is higher than the $R_{1}$ velocity, at 4,600 $\pm$ 200 km/s compared to 3,890 $\pm$ 50 km/s, respectively.  These faster initial velocities in the torus model imply that higher latitude radio emission played a large contribution during this period.  From Day 7,000 to 9,300, we see the Torus/$R_{1}$ ratio decrease as emission from the low-lying areas begins to dominate, which corresponds with the period when the torus velocity decreases to 2,300 $\pm$ 200 km/s and $R_{1}$ remains constant at 3,800 $\pm$ 460 km/s in our analysis.  

\begin{figure}
	\includegraphics[width=\columnwidth]{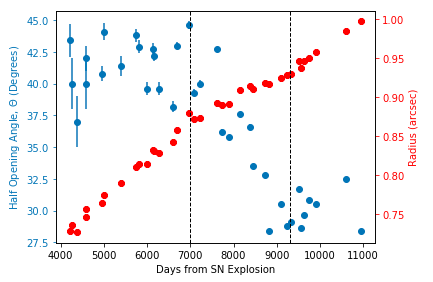}
    \caption{The measured half-opening angle ($\theta$, blue) in the SN 1987A remnant compared with the radii measured in the torus model (red).  Black dashed lines are provided for reference at the two changes in expansion seen in the torus radii at Day 7,000 and Day 9,300, respectively.}
    \label{fig:torusandtheta}
\end{figure}

The system re-stabilized at Day 9,300 before the ring and torus models could converge, meaning emission from high latitudes continues to be a factor in the radio remnant.  The existence of such emission would also help explain the expansion of both the torus and $R_{1}$ sizes beyond the optically observed ER before the shockwave interacted with it \citep{Plait1995}.  Further, there is a clear correlation between the expansion of the torus radii and $\theta$ during this time period, as shown in Figure \ref{fig:torusandtheta}.  

\begin{figure}
	\includegraphics[width=\columnwidth]{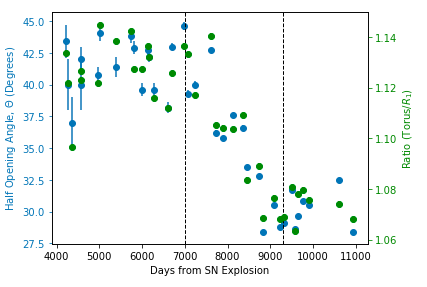}
    \caption{As for Figure \ref{fig:torusandtheta}, but with the measured half-opening angle plotted with the ratio between the torus model and $R_{1}$ (green).}
    \label{fig:torus$R_{1}$andtheta}
\end{figure}

We also see a coupling between the torus/$R_{1}$ radius ratio and the half opening angle, as shown in Figure \ref{fig:torus$R_{1}$andtheta}.  This implies that the radio emission has become more ring-like and two dimensional in nature over time, and that the emission is increasingly dominated by lower latitudes.  This could be due to one of two reasons.  First, it is possible that there previously was a higher amount of high-latitude emission, which has steadily decreased over time.  The second option is that while high-latitude emission is still present, the emission from lower latitudes has increasingly dominated, as the shockwave interacted with increasingly dense regions of the ER \citep{Potter2014}.

Finally, we should note that this occurred in conjunction with a decrease in the $R_{2}$/$R_{1}$ ratio from 0.78 $\pm$ 0.01 to 0.74 $\pm$ 0.02, respectively (Figure \ref{fig:R1byR2}).  The ER region is known to have an orientation of $41^{\circ}$ to the line of sight \citep{Sugerman2005}, which would imply a ratio of $\sim$0.70.  This supports the interpretation that some high latitude emission has been persistent, but that emission from the ER has dominated more recently.  

\subsubsection{Size Comparison to X-ray}
\label{sec:compare-to-x-ray}

The ring model allows us to compare the size of the remnant with X-ray data, which appears to follow a ring model fit \citep{Frank2016}.  When comparing the ring model to the X-ray data, the radio emitting region has consistently been larger than the X-ray emitting region since Day 7,500, which was first noted by \citet{Ng2013}.  Since this date, the trend has continued (Figure \ref{fig:radius-all}).  This difference is likely physical as opposed to a difference in measuring technique between X-ray and radio \citep{Gaensler2007,Ng2009,Ng2013}.

\citet{Ng2009} applied the torus spatial model to Day 7736 X-ray observations, and found a value of $\theta= 26^{\circ} \pm 3^{\circ}$.  We note that radio observations at this time yielded a value for $\theta$ of $36.2^{\circ}$-- that is, at the beginning of the transition of $\theta$ from a higher value to its lower one (Figure \ref{fig:theta}).  Further, the X-ray emission at this time is believed to originate from the shockwave interacting with the dense clumps in the ER \citep{Orlando2015,Frank2016}.  The fact that $\theta$ derived from our torus model from Day 8,800 is in agreement with the $\theta$ obtained by \citet{Ng2009} suggests the radio emission also largely originates from the same half opening angle region as is the X-ray radiation during this time, although the overall emission torus may be larger in radio.

\subsubsection{Asymmetry}
\label{sec:Asymmetry}

The morphological changes in SN 1987A have also included a reduction in the asymmetry of surface brightness between the eastern and western parts of the remnant (Figure \ref{fig:asymmetry}). A 3D simulation of SN 1987A by \citet{Potter2014} used this assessment, and predicted that the faster shocks would first depart the eastern lobe at Day 7,000, and shocks from the western lobe would later emerge at Day 8,000.  At this point, the model predicted that the asymmetry between the two lobes should pass parity (where the emission is equally bright between the eastern and western lobes), and ultimately the asymmetry should be brighter in the western side by Day 9,000-10,000.  However, we do not observe this in our data; instead, it appears that we are only beginning to approach parity between the eastern and western lobes (where both are equally bright) by Day 11,000.  This could be attributed to several possibilities, such as an over-density compared to what was used in theoretical models in the eastern lobe, which would delay the egress of the shock from this area.  Another possibility is that the assumed distribution in emission used by \citet{Potter2014} between the forward and reverse shocks is incorrect, and that there is instead a longer exit phase as different shocks leave the ring.  It should be noted, however, that despite the delay, our observations show a trend which is consistent with model predictions.

We can also compare the asymmetry to what we see in X-ray data.  Although it is not known whether the X-ray and radio emission necessarily originate from the same region, \citet{Frank2016} note that the radio emission evolves similarly to the X-ray and optical data, but is delayed by $\sim$2,000 days, which appears consistent with our findings (Figure \ref{fig:supernova-shots}).  They further suggest that the `hard' component ($\sim$2-10 keV) best matches the radio data in morphology, and that the east-west asymmetry only began to reverse in this X-ray band around Day $\sim$9,500.  This delay would be consistent with what we see in our data, where by Day 11,000 we have not yet reached parity in the brightness between the eastern and western lobes.  As such, it does appear that there is some similarity in the emission region from the ER between the radio and X-ray data.

\subsubsection{Flux Density}
\label{sec:flux-discussion}

Radio emission in SN 1987A is thought to be distributed between the forward and reverse shocks \citep{Jun1996}, although the exact distribution between these is still unknown.  In the case of SN 1987A, the picture is further complicated by potential emission from multiple components \citep{Blondin1996}, such as a component arising from the ER and emission from high latitudes.  Our residuals when the model was subtracted from the data favored the torus model over the ring model.  This fact suggests the picture including high-latitude emission is more accurate.

Radio emission in the period from Day 5,000 to 7,500 increases exponentially (Figure \ref{fig:flux-scatter}), and was thought to be caused by interactions between the shockwave and the ER \citep{Zanardo2010}.  From Day 7,500, we still see an increase in brightening of the remnant, but at a reduced rate compared to that in earlier epochs (Figure \ref{fig:flux-fit}).  This transition likely corresponds with a decrease of material interacting with the shockwave during this time, likely corresponding to the shockwave beginning to leave the ER.  If the asymmetry of the ER is caused by faster shocks in the east (see Section \ref{sec:Asymmetry}), as the shockwave leaves the eastern lobe a part of the shockwave would still be interacting with the western lobe, and this would cause the overall flux of the ER to continue increasing until the entire shockwave leaves the ring completely.

\subsubsection{Summary of Day 4,000-9,300}

The Day 4,000-9,300 period was clearly a time of transition in SN 1987A, with many changes observed over the epoch.  First, our observed expansion rate in both torus and ring models isconsistent with the shockwave interacting with the dense ER.  Second, there is a coupling between $\theta$ and the torus radii, as shown in Figure \ref{fig:torusandtheta}, whereby the decrease in the rate of expansion of the torus radii at Day 7,000 corresponds with the half-opening angle for the emission decreasing.  The expansion rate for the torus radii then begin to increase again once $\theta$ stabilizes.  We attribute this to high latitude emission being present from earlier stages, and then later the lower latitudes begining to dominate the emission profile.  The size of the radio remnant also appears to be larger than that of the X-ray remnant.  Further, the radio remnant appears to lag in flux density over time compared to what is seen in the X-ray by $\sim$2,000 days.

The asymmetry observed between the eastern and western limbs of the ER is delayed compared to predictions in theoretical models, which during this period predicted parity between brightness in both sides, and that the western limb would become subsequently brighter.  Instead, we see the eastern limb remaining brighter than the western limb.  Finally, the flux density of the SNR has continued to increase, but from Day 7,500, the increase has continued at a lower rate than what was measured for Days 4000-7500.  This could be due to the shockwave leaving the ring from this period, but the contribution of any emission from high latitudes that has been present since the ER was first encountered is unknown.

\subsection{Day 9,300 Onwards}
\label{sec:9300}

From Day 9,300, the supernova remnant has undergone several changes.  These include the re-acceleration to a faster expansion rate in the torus model fits to the data to 3,610 $\pm$ 240 km/s (Table \ref{tab:expansion} and Figure \ref{fig:radius-all}).  This is in agreement with the velocity value obtained for $R_{1}$, as well as $\theta$ stabilizing at 29.9 $\pm$ 0.1 degrees, as seen in Figure \ref{fig:torusandtheta}.  This re-acceleration has corresponded with an expansion rate $R \propto t^{m}$ of $m = 0.43 \pm$ 0.02 for the torus model, and 0.44 $\pm$ 0.02 for $R_{1}$.   This indicates that a transition occurred around Day 9,300, which we interpret as the forward shock leaving the ER.  

As the SN expands beyond the ER, high-latitude emission components of the radio remnant become more prominent (Potter et al. 2014). Given the hourglass structure of the nebula and its inclination along the line of sight, emission components above and below the equatorial plane have been identified as extended bright sites in the northern and southern sectors of the SNR \citep{Zanardo2018}. At 9 GHz, although we cannot distinguish the specific contribution of high-latitude emission from that of emission sites specifically located within the equatorial plane, i.e., where the dense CSM is located, we measured a marked increase of the emission from the northern and southern sites. In particular, within ~1,000 days (Day 9,915-10,942), the emission observed in the northern and southern sectors has undergone an increase of 20\%, compared to the 10\% increase across the eastern lobe (see flux density contours in Figure 2).  At day 11,000, the integrated flux density over the northern and southern sites is 12 mJy, while the emission from the brightest eastern sites is 18 mJy.  Further, \citet{Ng2013} predicted that if shocks from the lower latitudes continue to dominate the emission, the radius measurements between both the torus and ring measurements should converge.  With the re-acceleration of the torus expansion rate, we do not see that this will be the case in the future.

The southern part of the emission paints a more complex picture, as we can see by Day 10,600 that the southeastern area of the ring has begun to fade (Figure \ref{fig:contours}).  This is consistent with the break-up seen in the X-ray and optical data \citep{Fransson2015,Frank2016}.  Hotspots beyond this area have also been observed in the optical, interpreted as due to the direct interaction of the shockwave with gas beyond the ER \citep{Fransson2015}.

\subsection{Future Predictions}
\label{sec:future}

The SN 1987A remnant is transitioning into a new phase.  Much of what happens next in SN 1987A depends on the density of  the area beyond the visible emission region which the shockwave is now entering.  \citet{Mattila:2010} estimated a mass for the ring, but emphasized that this calculation only included the mass ionized by the initial shock breakout from the supernova.  As such, it is possible that the density beyond the visible ER could be greater than expected, assuming that the shockwave will enter a region where the CSM will originate from a free RSG wind emitted by the progenitor star \citep{Chevalier1995}.

Observations of hotspots appearing beyond the ER at optical wavelengths \citep{Fransson2015} also indicate that the structure beyond the ER may be more complex than expected.  Understanding this structure is particularly important if the CSM is from a binary progenitor model \citep{Urushibata2018, Menon2017, Morris2009}, as the distribution of material could shed light on slow and fast merger scenarios.  Radio observations should tell us more about this environment, as a further increase in shockwave velocity over the next few thousand days would indicate that the density of gas with which the shockwave is interacting is decreasing.  Simulations from \citet{Potter2014} predicted a velocity at a few thousand days as high as 6,000 km/s, although this is dependent on the temperature and density for the region.

When it comes to the asymmetry of the SN emission, if we extrapolate the rate of asymmetry decrease seen in Figure \ref{fig:asymmetry}, we expect the asymmetry to reach parity (where the east and west sides are equally bright), on Day 11,650 $\pm$ 60 using the torus model, and Day 12,300 $\pm$ 60 with the ring model.  This prediction is also consistent with the picture seen in X-rays, which is typically 2,000 days ahead of the radio \citep{Frank2016}.  We also expect this asymmetry to then increase as the western limb increases in brightness and the eastern limb continues to fade, although we note that the rate of asymmetry may change depending on the rate of destruction of the ER in the east, and on whether there are notable amounts of unionized or neutral gas beyond the ER which may be as dense as the ER, but not visible.  The picture on the western limb may also be more complex, as analysis by \citet{Zanardo:2014gu} suggested residual emission in the western region at higher frequencies that may be attributed to a pulsar wind nebula (PWN).

High latitude emission may also play a factor in the asymmetry, although we note that the difference in brightness between these two nodes matches very well the picture seen in X-ray data, which indicates that the two bright eastern and western nodes are likely originating from the same region.  Currently, the asymmetry values between the torus and ring models are in agreement, which also indicates that the emission creating the asymmetry comes from this region.  A divergence in the future asymmetry between the two models would indicate the presence of emission from higher latitudes.  Further, if there are notable amounts of polar emission emerging in the next few thousand days, we expect to see further brightening in both the northern and southern areas of the ER.  Such emission would be accompanied by an increase in $\theta$.

Finally, the radio flux for the SNR will provide a good indicator for any future high latitude emission beyond the ER.  Little is known of this region, although its contents are thought to be from the RSG wind of the progenitor \citep{Chevalier1995,Mattila:2010}.  If the material in the SNR is concentrated wholly in the ER, we expect the radio flux to plateau in the next few thousand days as the reverse shock leaves the ER, similar to what was seen in the X-ray emission \citep{Potter2014,Frank2016}.  New material interacting with the shockwaves, however, could contribute to a further increase in radio flux.

\subsection{Comparison to Other Supernovae}
\label{sec:compare}

SN 1987A provides a unique opportunity for comparisons between different radio supernovae, as its proximity means details in the structure can be seen which can only be resolved in a handful of other radio supernovae.  Although these other radio supernovae are brighter and have a different CSM density, these comparisons over time allow us to learn how common features such as asymmetry and changes in shockwave velocity are in young supernova environments.  In order to highlight these similarities, in this section we will compare our observations of SN 1987A with those of two other spatially resolved radio supernova remnants, SN 1986J and SN 1993J.

\subsubsection{SN 1986 J}

SN 1986J was first detected in 1986 via its radio emission, although it is estimated that its explosion epoch is 1983.2 $\pm$1.1 \citep{Rupen1987,Bietenholz2002}.  It is a Type IIn supernova located at a distance of $\sim$10 Mpc \citep{Rupen1987}.  Radio emission from the expanding shell of the SNR was the dominant observed emission for many years \citep{Bietenholz2010}, but at present the emission of the shell is decreasing, with the brightness of the outer edge fading faster than the edge near the center \citep{Bietenholz:2017bf}.  This is because SN 1986J also has an observable central component at its center, first detected at t $\simeq$ 20 years, which is now approximately ten times brighter than the shell component.

Radio observations of the SN 1986J shell have included a hotspot region brighter than the rest of the shell once resolution was sufficient to observe it \citep{Bietenholz2002}.  This is consistent with other spatially resolved radio supernovae, such as SN 1993J \citep{Bietenholz2010}, SN 2011dh \citep{deWitt2016}, and SN 2014C \citep{Bietenholz2017b}, where the brighter hotspot regions have expanded homologously with the SNR as the shockwave expanded through the surrounding regions.  Further, the SN 1986J shell was reported by \citet{Bietenholz:2017bf} as expanding at a velocity of 2,810 $\pm$ 750 km/s, which is consistent with the velocities observed in SN 1987A.  As such, it appears that the expansion in both remnants is dominated by the forward shock of the supernova event itself.  However, it is likely that the CSM surrounding SN 1986J is much more dense than that of SN 1987A based on the former's optical spectral lines \citep{Filippenko1997}.

Overall, the expansion of the ring, and the expansion of homologous hotspots with this area of the remnant in SN 1986J are consistent with our observations of SN 1987A.  However, the CSM surrounding SN 1986J is distributed very differently, as evidenced by how radio luminosity was likely created by material in the CSM and that the ring is now fading.  In the future of SN 1987A, however, any potential increase in radio luminosity detected on the inner edge of the observed ring may be evidence of a compact object.  If the compact object in SN 1987A is off-center in the SNR, as suggested by \citet{Zanardo:2014gu}, such a brightening would most likely be visible on the inner western lobe.

\subsubsection{SN 1993J}

SN 1993J, at a distance of 3.62 Mpc in M81, is the second brightest supernova observed in the last century after SN 1987A.  A Type IIn supernova, its progenitor is believed to have had a significantly different mass loss history compared to SN 1987A, with a simple $\rho \propto r^{-2}$ CSM during its first few hundred days \citep{Weiler2007,SS1993}.  A radio shell was first observed 175 days after the explosion using Very Long Baseline Interferometry (VLBI) imaging, with multiple hotspots that shift in orientation through 2,787 days \citep{Bietenholz2010}.  

In the SN 1993J disc, an asymmetry in brightness is first seen when the disc appears, in the south-east, but this hotspot appears to shift within a few hundred days and disappear altogether by $\sim$1,000 days after the supernova explosion \citep{Bietenholz2003}.  This is much faster than what we observe for hotspots in SN 1987A, and may imply greater density disparities in the SN 1993J ejecta than that seen in the ER for SN 1987A.  Further, asymmetry has been suggested for the SN 1993J explosion based on optical spectra \citep{Lewis1994}, which may also cause radio hotspots similar to those observed for SN 1987A.

\citet{Bietenholz2001} fit the observed SN 1993J remnant with an optically thin spherical shell model, which they felt was accurate as the radio emission was circular to within 4\%.  If we compare our $R_{2}$/$R_{1}$ ratio for a two dimensional model of the emission ring, where we expect a 0.70 ratio based on the inclination of the system, we find that the emission from SN 1987A was more asymmetric during the period when the shockwave was interacting with the ER.  This value decreased, however, to $0.74\pm0.02$ in recent observations, meaning that the emission has become more circular over time.  This shows that the CSM emitted by the SN 1993J progenitor was more uniform than the complex structure surrounding SN 1987A.

\section{Conclusions}

We have reported our imaging results of SN 1987A at 9 GHz using ATCA, covering a 25-year period from 1992 to 2017.  We have also carried out Fourier modeling of the visibilities, with a torus model and ring model used to describe the evolving remnant.  Both models have shown the continued expansion of SN 1987A through February 2017, as well as an increasing flux density.  Our data are most consistent with a torus model where high latitude emission is present from soon after the shockwave encountered the ER.  This is based on the rate of expansion of the remnant, the observed changes in the evolution of the torus radius at Day 7,000 and Day 9,300, and how this data is also coupled with the decrease of the half-opening angle of the remnant during this period.  Lower latitude emission then dominates during the later stages as the shockwave continues to plow through the ER.  We should note that the radio remnant appears to be larger than the X-ray remnant, although the radio remnant appears to lag 2,000 days after the morphology seen in X-ray data, which may be due to the magnetic fields in the remnant increasing in strength after the shockwave has passed through the medium.

Unlike at other wavelengths, we have not yet seen the western side of the radio remnant become brighter than the eastern side.  We have also seen the southeastern side of the SNR begin to fade, which, combined with the similar fading seen in X-ray and optical data, suggests that the shockwave has left at least this region of the ER.

In the future, we expect the western side of the SNR to become the brightest region as the eastern continues to fade, and we also expect a further plateau in radio emission as the shockwave leaves the ER completely.  Our observations will also help us understand the structure of material beyond the ER, which is from the progenitor star's stellar wind and about which little is known.  Because of its proximity, studies of SN 1987A will also be useful for the comparison to other radio supernovae and their surrounding CSM.

\section{Acknowledgements}

We would like to thank the referee for their helpful comments in the preparation of this manuscript.  We thank K. Frank for recent Chandra data.  The Australia Telescope Compact Array is part of the Australia Telescope, which is funded by the Commonwealth of Australia for operation as a National
Facility managed by CSIRO.  The Dunlap Institute is funded through an endowment established by the David Dunlap family and the University of Toronto. B.M.G. acknowledges the support of the Natural Sciences and Engineering Research Council of Canada (NSERC) through grant RGPIN-2015-05948, and of the Canada Research Chairs program.

\facilities{ATCA}

%\facility{facility ID}
\software{MIRIAD, SciPy}

\bibliographystyle{yahapj}
\bibliography{sn1987a}

\begin{thebibliography}{}
\providecommand\natexlab[1]{#1}
\providecommand\JournalTitle[1]{#1}

\bibitem[{{Ball} {et~al.}(1995){Ball}, {Campbell-Wilson}, {Crawford}, \&
  {Turtle}}]{Ball1995}
{Ball}, L., {Campbell-Wilson}, D., {Crawford}, D.~F., \& {Turtle}, A.~J. 1995,
  \href{http://dx.doi.org/10.1086/176446}{\JournalTitle{\apj}, 453, 864}

\bibitem[{{Ball} \& {Kirk}(1992)}]{Ball1992}
{Ball}, L., \& {Kirk}, J.~G. 1992,
  \href{http://dx.doi.org/10.1086/186512}{\JournalTitle{\apjl}, 396, L39}

\bibitem[{Bietenholz \& Bartel(2017)}]{Bietenholz:2017bf}
Bietenholz, M.~F., \& Bartel, N. 2017, \JournalTitle{ApJ}, 839, 10

\bibitem[{{Bietenholz} {et~al.}(2001){Bietenholz}, {Bartel}, \&
  {Rupen}}]{Bietenholz2001}
{Bietenholz}, M.~F., {Bartel}, N., \& {Rupen}, M.~P. 2001,
  \href{http://dx.doi.org/10.1086/321647}{\JournalTitle{\apj}, 557, 770}

\bibitem[{{Bietenholz} {et~al.}(2002){Bietenholz}, {Bartel}, \&
  {Rupen}}]{Bietenholz2002}
---. 2002, \href{http://dx.doi.org/10.1086/344251}{\JournalTitle{\apj}, 581,
  1132}

\bibitem[{{Bietenholz} {et~al.}(2003){Bietenholz}, {Bartel}, \&
  {Rupen}}]{Bietenholz2003}
---. 2003, \href{http://dx.doi.org/10.1086/378265}{\JournalTitle{\apj}, 597,
  374}

\bibitem[{{Bietenholz} {et~al.}(2010){Bietenholz}, {Bartel}, \&
  {Rupen}}]{Bietenholz2010}
---. 2010,
  \href{http://dx.doi.org/10.1088/0004-637X/712/2/1057}{\JournalTitle{\apj},
  712, 1057}

\bibitem[{{Bietenholz} {et~al.}(2018){Bietenholz}, {Kamble}, {Margutti},
  {Milisavljevic}, \& {Soderberg}}]{Bietenholz2017b}
{Bietenholz}, M.~F., {Kamble}, A., {Margutti}, R., {Milisavljevic}, D., \&
  {Soderberg}, A. 2018,
  \href{http://dx.doi.org/10.1093/mnras/stx3194}{\JournalTitle{\mnras}, 475,
  1756}

\bibitem[{{Blondin} {et~al.}(1996){Blondin}, {Lundqvist}, \&
  {Chevalier}}]{Blondin1996}
{Blondin}, J.~M., {Lundqvist}, P., \& {Chevalier}, R.~A. 1996,
  \href{http://dx.doi.org/10.1086/178060}{\JournalTitle{\apj}, 472, 257}

\bibitem[{{Briggs}(1994)}]{Briggs1994}
{Briggs}, D.~S. 1994, in The Restoration of HST Images and Spectra - II, ed.
  R.~J. {Hanisch} \& R.~L. {White}, 250

\bibitem[{{Briggs}(1995)}]{Briggs1995}
{Briggs}, D.~S. 1995, in BAAS, Vol.~27, BAAS, 1444

\bibitem[{{Burrows} {et~al.}(1995){Burrows}, {Krist}, {Hester}, {Sahai},
  {Trauger}, {Stapelfeldt}, {Gallagher}, {Ballester}, {Casertano}, {Clarke},
  {Crisp}, {Evans}, {Griffiths}, {Hoessel}, {Holtzman}, {Mould}, {Scowen},
  {Watson}, \& {Westphal}}]{Burrows1995}
{Burrows}, C.~J., {Krist}, J., {Hester}, J.~J., {et~al.} 1995,
  \href{http://dx.doi.org/10.1086/176339}{\JournalTitle{\apj}, 452, 680}

\bibitem[{{Chevalier}(1982)}]{Chevalier1982B}
{Chevalier}, R.~A. 1982,
  \href{http://dx.doi.org/10.1086/160167}{\JournalTitle{\apj}, 259, 302}

\bibitem[{{Chevalier} \& {Dwarkadas}(1995)}]{Chevalier1995}
{Chevalier}, R.~A., \& {Dwarkadas}, V.~V. 1995,
  \href{http://dx.doi.org/10.1086/309714}{\JournalTitle{\apjl}, 452, L45}

\bibitem[{{Chevalier} \& {Imamura}(1982)}]{Chevalier1982}
{Chevalier}, R.~A., \& {Imamura}, J.~N. 1982,
  \href{http://dx.doi.org/10.1086/160364}{\JournalTitle{\apj}, 261, 543}

\bibitem[{{Chita} {et~al.}(2008){Chita}, {Langer}, {van Marle},
  {Garc{\'{\i}}a-Segura}, \& {Heger}}]{Chita2008}
{Chita}, S.~M., {Langer}, N., {van Marle}, A.~J., {Garc{\'{\i}}a-Segura}, G.,
  \& {Heger}, A. 2008,
  \href{http://dx.doi.org/10.1051/0004-6361:200810087}{\JournalTitle{\aap},
  488, L37}

\bibitem[{{de Witt} {et~al.}(2016){de Witt}, {Bietenholz}, {Kamble},
  {Soderberg}, {Brunthaler}, {Zauderer}, {Bartel}, \& {Rupen}}]{deWitt2016}
{de Witt}, A., {Bietenholz}, M.~F., {Kamble}, A., {et~al.} 2016,
  \href{http://dx.doi.org/10.1093/mnras/stv2306}{\JournalTitle{\mnras}, 455,
  511}

\bibitem[{{Filippenko}(1997)}]{Filippenko1997}
{Filippenko}, A.~V. 1997,
  \href{http://dx.doi.org/10.1146/annurev.astro.35.1.309}{\JournalTitle{\araa},
  35, 309}

\bibitem[{{Frank} {et~al.}(2016){Frank}, {Zhekov}, {Park}, {McCray}, {Dwek}, \&
  {Burrows}}]{Frank2016}
{Frank}, K.~A., {Zhekov}, S.~A., {Park}, S., {et~al.} 2016,
  \href{http://dx.doi.org/10.3847/0004-637X/829/1/40}{\JournalTitle{\apj}, 829,
  40}

\bibitem[{{Fransson} {et~al.}(2015){Fransson}, {Larsson}, {Migotto}, {Pesce},
  {Challis}, {Chevalier}, {France}, {Kirshner}, {Leibundgut}, {Lundqvist},
  {McCray}, {Spyromilio}, {Taddia}, {Jerkstrand}, {Mattila}, {Smith},
  {Sollerman}, {Wheeler}, {Crotts}, {Garnavich}, {Heng}, {Lawrence}, {Panagia},
  {Pun}, {Sonneborn}, \& {Sugerman}}]{Fransson2015}
{Fransson}, C., {Larsson}, J., {Migotto}, K., {et~al.} 2015,
  \href{http://dx.doi.org/10.1088/2041-8205/806/1/L19}{\JournalTitle{\apjl},
  806, L19}

\bibitem[{{Gaensler} {et~al.}(1997){Gaensler}, {Manchester}, {Staveley-Smith},
  {Tzioumis}, {Reynolds}, \& {Kesteven}}]{Gaensler1997}
{Gaensler}, B.~M., {Manchester}, R.~N., {Staveley-Smith}, L., {et~al.} 1997,
  \href{http://dx.doi.org/10.1086/303917}{\JournalTitle{\apj}, 479, 845}

\bibitem[{{Gaensler} {et~al.}(2007){Gaensler}, {Staveley-Smith}, {Manchester},
  {Kesteven}, {Ball}, \& {Tzioumis}}]{Gaensler2007}
{Gaensler}, B.~M., {Staveley-Smith}, L., {Manchester}, R.~N., {et~al.} 2007,
  \href{http://dx.doi.org/10.1063/1.3682887}{in American Institute of Physics
  Conference Series, Vol. 937, Supernova 1987A: 20 Years After: Supernovae and
  Gamma-Ray Bursters, ed. S.~{Immler}, K.~{Weiler}, \& R.~{McCray}}, 86

\bibitem[{{Gull} \& {Daniell}(1978)}]{GD1978}
{Gull}, S.~F., \& {Daniell}, G.~J. 1978,
  \href{http://dx.doi.org/10.1038/272686a0}{\JournalTitle{\nat}, 272, 686}

\bibitem[{{Helder} {et~al.}(2013){Helder}, {Broos}, {Dewey}, {Dwek}, {McCray},
  {Park}, {Racusin}, {Zhekov}, \& {Burrows}}]{Helder2013}
{Helder}, E.~A., {Broos}, P.~S., {Dewey}, D., {et~al.} 2013,
  \href{http://dx.doi.org/10.1088/0004-637X/764/1/11}{\JournalTitle{\apj}, 764,
  11}

\bibitem[{{Hogg} {et~al.}(2010){Hogg}, {Bovy}, \& {Lang}}]{Hogg2010}
{Hogg}, D.~W., {Bovy}, J., \& {Lang}, D. 2010, \JournalTitle{ArXiv e-prints},
  \href{http://arxiv.org/abs/1008.4686}{{\sffamily arXiv:1008.4686
  [astro-ph.IM]}}

\bibitem[{Jones {et~al.}(2001)Jones, Oliphant, Peterson, {et~al.}}]{SciPy}
Jones, E., Oliphant, T., Peterson, P., {et~al.} 2001, {SciPy}: Open source
  scientific tools for {Python}, [Online; accessed May 2, 2018]

\bibitem[{{Jun} \& {Norman}(1996)}]{Jun1996}
{Jun}, B.-I., \& {Norman}, M.~L. 1996,
  \href{http://dx.doi.org/10.1086/178059}{\JournalTitle{\apj}, 472, 245}

\bibitem[{Kass \& Raftery(1995)}]{Kass1995}
Kass, R.~E., \& Raftery, A.~E. 1995,
  \href{http://dx.doi.org/10.1080/01621459.1995.10476572}{\JournalTitle{Journal
  of the American Statistical Association}, 90, 773}

\bibitem[{{Lewis} {et~al.}(1994){Lewis}, {Walton}, {Meikle}, {Martin},
  {Cumming}, {Catchpole}, {Arevalo}, {Argyle}, {Benn}, {Bunclark}, {Castaneda},
  {Centurion}, {Clegg}, {Delgado}, {Dhillon}, {Goudfrooij}, {Harlaftis},
  {Hassall}, {Helmer}, {Hill}, {Jones}, {King}, {Lazaro}, {Lucey}, {Martin},
  {Miller}, {Morrison}, {Penny}, {Perez}, {Read}, {Rudd}, {Rutten}, {Sharples},
  {Unger}, \& {Vilchez}}]{Lewis1994}
{Lewis}, J.~R., {Walton}, N.~A., {Meikle}, W.~P.~S., {et~al.} 1994,
  \href{http://dx.doi.org/10.1093/mnras/266.1.L27}{\JournalTitle{\mnras}, 266,
  L27}

\bibitem[{Mattila {et~al.}(2010)Mattila, Lundqvist, Gr{\"o}ningsson, Meikle,
  Stathakis, Fransson, \& Cannon}]{Mattila:2010}
Mattila, S., Lundqvist, P., Gr{\"o}ningsson, P., {et~al.} 2010,
  \JournalTitle{ApJ}, 717, 1140

\bibitem[{{McCray} \& {Fransson}(2016)}]{McCray2016}
{McCray}, R., \& {Fransson}, C. 2016,
  \href{http://dx.doi.org/10.1146/annurev-astro-082615-105405}{\JournalTitle{\araa},
  54, 19}

\bibitem[{{Menon} \& {Heger}(2017)}]{Menon2017}
{Menon}, A., \& {Heger}, A. 2017,
  \href{http://dx.doi.org/10.1017/S1743921317003003}{in IAU Symposium, Vol.
  329, The Lives and Death-Throes of Massive Stars, ed. J.~J. {Eldridge}, J.~C.
  {Bray}, L.~A.~S. {McClelland}, \& L.~{Xiao}}, 64

\bibitem[{{Morris} \& {Podsiadlowski}(2007)}]{Morris2007}
{Morris}, T., \& {Podsiadlowski}, P. 2007,
  \href{http://dx.doi.org/10.1126/science.1136351}{\JournalTitle{Science}, 315,
  1103}

\bibitem[{{Morris} \& {Podsiadlowski}(2009)}]{Morris2009}
---. 2009,
  \href{http://dx.doi.org/10.1111/j.1365-2966.2009.15114.x}{\JournalTitle{\mnras},
  399, 515}

\bibitem[{{Ng} {et~al.}(2009){Ng}, {Gaensler}, {Murray}, {Slane}, {Park},
  {Staveley-Smith}, {Manchester}, \& {Burrows}}]{Ng2009}
{Ng}, C.-Y., {Gaensler}, B.~M., {Murray}, S.~S., {et~al.} 2009,
  \href{http://dx.doi.org/10.1088/0004-637X/706/1/L100}{\JournalTitle{\apjl},
  706, L100}

\bibitem[{{Ng} {et~al.}(2008){Ng}, {Gaensler}, {Staveley-Smith}, {Manchester},
  {Kesteven}, {Ball}, \& {Tzioumis}}]{Ng2008}
{Ng}, C.-Y., {Gaensler}, B.~M., {Staveley-Smith}, L., {et~al.} 2008,
  \href{http://dx.doi.org/10.1086/590330}{\JournalTitle{\apj}, 684, 481}

\bibitem[{{Ng} {et~al.}(2013){Ng}, {Zanardo}, {Potter}, {Staveley-Smith},
  {Gaensler}, {Manchester}, \& {Tzioumis}}]{Ng2013}
{Ng}, C.-Y., {Zanardo}, G., {Potter}, T.~M., {et~al.} 2013,
  \href{http://dx.doi.org/10.1088/0004-637X/777/2/131}{\JournalTitle{\apj},
  777, 131}

\bibitem[{{Orlando} {et~al.}(2015){Orlando}, {Miceli}, {Pumo}, \&
  {Bocchino}}]{Orlando2015}
{Orlando}, S., {Miceli}, M., {Pumo}, M.~L., \& {Bocchino}, F. 2015,
  \href{http://dx.doi.org/10.1088/0004-637X/810/2/168}{\JournalTitle{\apj},
  810, 168}

\bibitem[{{Plait} {et~al.}(1995){Plait}, {Lundqvist}, {Chevalier}, \&
  {Kirshner}}]{Plait1995}
{Plait}, P.~C., {Lundqvist}, P., {Chevalier}, R.~A., \& {Kirshner}, R.~P. 1995,
  \href{http://dx.doi.org/10.1086/175213}{\JournalTitle{\apj}, 439, 730}

\bibitem[{Potter {et~al.}(2014)Potter, Staveley-Smith, Reville, Ng, Bicknell,
  Sutherland, \& Wagner}]{Potter2014}
Potter, T.~M., Staveley-Smith, L., Reville, B., {et~al.} 2014,
  \JournalTitle{ApJ}, 794, 174

\bibitem[{{Racusin} {et~al.}(2009){Racusin}, {Park}, {Zhekov}, {Burrows},
  {Garmire}, \& {McCray}}]{Racusin2009}
{Racusin}, J.~L., {Park}, S., {Zhekov}, S., {et~al.} 2009,
  \href{http://dx.doi.org/10.1088/0004-637X/703/2/1752}{\JournalTitle{\apj},
  703, 1752}

\bibitem[{{Rupen} {et~al.}(1987){Rupen}, {van Gorkom}, {Knapp}, {Gunn}, \&
  {Schneider}}]{Rupen1987}
{Rupen}, M.~P., {van Gorkom}, J.~H., {Knapp}, G.~R., {Gunn}, J.~E., \&
  {Schneider}, D.~P. 1987,
  \href{http://dx.doi.org/10.1086/114447}{\JournalTitle{\aj}, 94, 61}

\bibitem[{{Sault} {et~al.}(1995){Sault}, {Teuben}, \& {Wright}}]{Sault1995}
{Sault}, R.~J., {Teuben}, P.~J., \& {Wright}, M.~C.~H. 1995, in Astronomical
  Society of the Pacific Conference Series, Vol.~77, Astronomical Data Analysis
  Software and Systems IV, ed. R.~A. {Shaw}, H.~E. {Payne}, \& J.~J.~E.
  {Hayes}, 433

\bibitem[{{Staveley-Smith} {et~al.}(1993){Staveley-Smith}, {Briggs}, {Rowe},
  {Manchester}, {Reynolds}, {Tzioumis}, \& {Kesteven}}]{SS1993}
{Staveley-Smith}, L., {Briggs}, D.~S., {Rowe}, A.~C.~H., {et~al.} 1993,
  \href{http://dx.doi.org/10.1038/366136a0}{\JournalTitle{\nat}, 366, 136}

\bibitem[{{Staveley-Smith} {et~al.}(1992){Staveley-Smith}, {Manchester},
  {Kesteven}, {Reynolds}, {Tzioumis}, {Killeen}, {Jauncey}, {Campbell-Wilson},
  {Crawford}, \& {Turtle}}]{SS1992}
{Staveley-Smith}, L., {Manchester}, R.~N., {Kesteven}, M.~J., {et~al.} 1992,
  \href{http://dx.doi.org/10.1038/355147a0}{\JournalTitle{\nat}, 355, 147}

\bibitem[{{Sugerman} {et~al.}(2005){Sugerman}, {Crotts}, {Kunkel}, {Heathcote},
  \& {Lawrence}}]{Sugerman2005}
{Sugerman}, B.~E.~K., {Crotts}, A.~P.~S., {Kunkel}, W.~E., {Heathcote}, S.~R.,
  \& {Lawrence}, S.~S. 2005,
  \href{http://dx.doi.org/10.1086/430408}{\JournalTitle{\apjs}, 159, 60}

\bibitem[{{Turtle} {et~al.}(1987){Turtle}, {Campbell-Wilson}, {Bunton},
  {Jauncey}, {Kesteven}, {Manchester}, {Norris}, {Storey}, \&
  {Reynolds}}]{Turtle1987}
{Turtle}, A.~J., {Campbell-Wilson}, D., {Bunton}, J.~D., {et~al.} 1987,
  \href{http://dx.doi.org/10.1038/327038a0}{\JournalTitle{\nat}, 327, 38}

\bibitem[{{Urushibata} {et~al.}(2018){Urushibata}, {Takahashi}, {Umeda}, \&
  {Yoshida}}]{Urushibata2018}
{Urushibata}, T., {Takahashi}, K., {Umeda}, H., \& {Yoshida}, T. 2018,
  \href{http://dx.doi.org/10.1093/mnrasl/slx166}{\JournalTitle{\mnras}, 473,
  L101}

\bibitem[{Weiler {et~al.}(2002)Weiler, Panagia, Montes, \& Sramek}]{Weiler2002}
Weiler, K.~W., Panagia, N., Montes, M.~J., \& Sramek, R.~A. 2002,
  \JournalTitle{Annual Review of Astronomy and Astrophysics}, 40, 387

\bibitem[{{Weiler} {et~al.}(2007){Weiler}, {Williams}, {Panagia}, {Stockdale},
  {Kelley}, {Sramek}, {Van Dyk}, \& {Marcaide}}]{Weiler2007}
{Weiler}, K.~W., {Williams}, C.~L., {Panagia}, N., {et~al.} 2007,
  \href{http://dx.doi.org/10.1086/523258}{\JournalTitle{\apj}, 671, 1959}

\bibitem[{{Wilson} {et~al.}(2011){Wilson}, {Ferris}, {Axtens}, {Brown},
  {Davis}, {Hampson}, {Leach}, {Roberts}, {Saunders}, {Koribalski}, {Caswell},
  {Lenc}, {Stevens}, {Voronkov}, {Wieringa}, {Brooks}, {Edwards}, {Ekers},
  {Emonts}, {Hindson}, {Johnston}, {Maddison}, {Mahony}, {Malu}, {Massardi},
  {Mao}, {McConnell}, {Norris}, {Schnitzeler}, {Subrahmanyan}, {Urquhart},
  {Thompson}, \& {Wark}}]{Wilson2011}
{Wilson}, W.~E., {Ferris}, R.~H., {Axtens}, P., {et~al.} 2011,
  \href{http://dx.doi.org/10.1111/j.1365-2966.2011.19054.x}{\JournalTitle{\mnras},
  416, 832}

\bibitem[{{Zanardo} {et~al.}(2018){Zanardo}, {Staveley-Smith}, {Gaensler},
  {Indebetouw}, {Ng}, {Matsuura}, \& {Tzioumis}}]{Zanardo2018}
{Zanardo}, G., {Staveley-Smith}, L., {Gaensler}, B.~M., {et~al.} 2018,
  \JournalTitle{ArXiv e-prints},
  \href{http://arxiv.org/abs/1806.04741}{{\sffamily arXiv:1806.04741
  [astro-ph.HE]}}

\bibitem[{{Zanardo} {et~al.}(2013){Zanardo}, {Staveley-Smith}, {Ng},
  {Gaensler}, {Potter}, {Manchester}, \& {Tzioumis}}]{Zanardo2013}
{Zanardo}, G., {Staveley-Smith}, L., {Ng}, C.-Y., {et~al.} 2013,
  \href{http://dx.doi.org/10.1088/0004-637X/767/2/98}{\JournalTitle{\apj}, 767,
  98}

\bibitem[{{Zanardo} {et~al.}(2010){Zanardo}, {Staveley-Smith}, {Ball},
  {Gaensler}, {Kesteven}, {Manchester}, {Ng}, {Tzioumis}, \&
  {Potter}}]{Zanardo2010}
{Zanardo}, G., {Staveley-Smith}, L., {Ball}, L., {et~al.} 2010,
  \href{http://dx.doi.org/10.1088/0004-637X/710/2/1515}{\JournalTitle{\apj},
  710, 1515}

\bibitem[{Zanardo {et~al.}(2014)Zanardo, Staveley-Smith, Indebetouw, Chevalier,
  Matsuura, Gaensler, Barlow, Fransson, Manchester, Baes, Kamenetzky, Laki~evi,
  Lundqvist, Marcaide, Mart~Vidal, Meixner, Ng, Park, Sonneborn, Spyromilio, \&
  van Loon}]{Zanardo:2014gu}
Zanardo, G., Staveley-Smith, L., Indebetouw, R., {et~al.} 2014,
  \JournalTitle{ApJ}, 796, 82

\end{thebibliography}

%\appendix
%\section{Appendix for Observations and Obtained Parameters- Torus and Ring Models}
%\label{sec:appendix}

\(\)

\end{document}